\documentclass[letterpaper]{article} 
\usepackage{aaai2027}
\usepackage{amsmath,amssymb,amsthm}

\usepackage[hyphens]{url} 
\usepackage{graphicx} 
\usepackage{natbib} 
\usepackage{caption} 
\usepackage{multirow}
\usepackage{algorithm}
\usepackage{algorithmic}

\usepackage{booktabs}
\usepackage{array}
\usepackage{tabularx}
\usepackage{colortbl}
\usepackage{xspace}

\newcommand{\methodname}{\textbf{\texttt{PROSE}}\xspace}

\usepackage{newfloat}
\usepackage{listings}

\DeclareCaptionStyle{ruled}{
  labelfont=normalfont,
  labelsep=colon,
  strut=off
} 

\floatstyle{ruled}
\newfloat{listing}{tb}{lst}{}
\floatname{listing}{Listing}

\definecolor{successgreen}{RGB}{205,245,200}
\definecolor{failred}{RGB}{250,205,205}

\newcommand{\proseyes}[1]{\cellcolor{green!15}#1}
\newcommand{\proseno}[1]{\cellcolor{red!12}#1}

\newcommand{\yes}{\cellcolor{successgreen}1}
\newcommand{\no}{\cellcolor{failred}0}
\newcommand{\yesb}{\cellcolor{successgreen}\textbf{1}}
\newcommand{\nob}{\cellcolor{failred}\textbf{0}}

\newcommand{\pvalyes}[1]{\cellcolor{green!15}#1}
\newcommand{\pvalno}[1]{\cellcolor{red!12}#1}

\newcommand{\fsryes}[1]{\cellcolor{green!15}#1}
\newcommand{\fsrno}[1]{\cellcolor{red!12}#1}

\title{The Shape of Ownership: Verifying LLM Provenance through Semantic Structures}

\author{
    Zhongrui Sun\textsuperscript{\rm 1}\equalcontrib, Jiahao Chen\textsuperscript{\rm 2}\equalcontrib, Oubo Ma\textsuperscript{\rm 2}, Yuwen Pu\textsuperscript{\rm 1}, \\ Zhou Feng\textsuperscript{\rm 2}, Haibo Hu\textsuperscript{\rm 1}, Shouling Ji\textsuperscript{\rm 2}\\
}
\affiliations{
  \textsuperscript{\rm 1} School of Big Data \& Software Engineering, Chongqing University\\
  \textsuperscript{\rm 2} College of Computer Science and Technology, Zhejiang University\\
}

\begin{document}

\maketitle

\begin{abstract}
As large language models (LLMs) are increasingly redistributed, adapted, and served through opaque APIs, model ownership can no longer be reliably established by inspecting model internals or deployment records. This creates a need for behavioral signatures that remain observable through black-box interaction. Yet most existing black-box fingerprints instantiate ownership signals through fixed query-key associations. They reduce model identity to sparse memorized associations detached from ordinary behavior, limiting both robustness and stealth. Therefore, a stronger fingerprint could instead be distributed, naturally elicited, and expressed at a higher semantic level. To this end, we introduce \methodname (\textbf{P}rovenance through \textbf{R}elational \textbf{O}rganization of \textbf{S}emantic \textbf{E}xpression), replacing fixed query sets with a semantical target domain and brittle response keys with semantic structures (i.e., abstract meaning representation) internalized as domain-conditioned response behavior. Specifically, the fingerprint is encoded in how the model semantically organizes its in-domain conclusions, rather than prescribed outputs. \methodname constructs a private bank of domain-specific semantic templates, internalizes them through mixed fine-tuning on structurally verified and clean responses, and verifies ownership by detecting the designated structures in responses to held-out natural queries. Extensive experiments across multiple model architectures, scales, and target domains show that \methodname achieves a 100\% fingerprint detection rate with no false positives, preserves model utility, and retains strong detectability under downstream modifications and output transformations.
\end{abstract}

\section{Introduction}
\label{sec:introduction}

Large language models (LLMs) are increasingly released as open-weight checkpoints, adapted into downstream variants, and deployed behind opaque APIs. After fine-tuning, compression, quantization, pruning, or knowledge distillation, a derived model may differ substantially from its source in parameters and external form while retaining much of its behavior~\cite{zhang2024reefrepresentationencodingfingerprints}. When weights, training records, and deployment logs are unavailable, model lineage must therefore be established solely through observable input-output behavior. This motivates \emph{black-box model fingerprinting}: embedding a private behavioral signature before release and later verifying whether a suspicious model has inherited it through textual interaction alone~\cite{xu-etal-2024-instructional,
russinovich2026heythatsmodelintroducing,
nasery2025scalablefingerprintinglargelanguage}.

Most existing black-box fingerprints encode ownership through a set of private query-response associations~\cite{xu2026copyrightprotectionlargelanguage}. Selected queries are mapped to predefined keys, rare strings, or designated responses, and ownership is verified by reproducing these associations~\cite{xu-etal-2024-instructional,
russinovich2026heythatsmodelintroducing,
nasery2025scalablefingerprintinglargelanguage,
xu2025ctccrobuststealthyfingerprinting,xu2025evertracer}. This design ties both fingerprint activation and evidence to sparse surface forms. Query rewriting, contextual changes, downstream adaptation, or output filtering may disrupt the associations, while anomalous queries and fixed responses can expose the verification intent~\cite{gloaguen2026llm}. Moreover, such methods represent model identity as isolated memorization rather than a persistent property of model behavior.


Recent works move beyond exact query-response memorization by broadening fingerprint activation to semantic input conditions or embedding signals above the token level~\cite{
gloaguen2026llm,ye2026swansemanticwatermarkingabstract}. However, the verification object remains largely unchanged: ownership is still inferred from whether individual outputs contain a designated artifact. This output-centric view does not capture a stronger form of model identity, a persistent behavioral regularity that emerges across different responses under the same semantic condition. We therefore ask: \emph{can ownership be encoded in the conditional distribution of how a model organizes meaning?} Answering this question requires a fingerprint that is naturally elicited by unseen in-domain queries, invariant to lexical and syntactic variation, selective against non-target behavior, and persistent under modification.

To this end, we propose \textbf{\methodname} (\textbf{P}rovenance through \textbf{R}elational \textbf{O}rganization of \textbf{S}emantic \textbf{E}xpression), a domain-conditioned semantic fingerprinting framework for black-box LLM ownership verification. \methodname treats ownership as a distributional property of semantic expression within a target domain. It replaces fixed private queries with natural target-domain queries, fixed response keys with private predicate-argument structures, Abstract Meaning Representation (AMR), and isolated exact matches with aggregated structural evidence across multiple interactions. The same fingerprint can therefore be expressed through different words, syntactic constructions, entities, and numerical values while preserving underlying semantic organization.

Specifically, \methodname instantiates these structures using domain-specific AMR templates. The owner first constructs a private template bank and uses a teacher LLM to generate target-domain responses that realize the designated structures. AMR parsing, semantic abstraction, and structural matching retain only verified responses, which are then mixed with clean instruction data to fine-tune the base model. After training, natural in-domain queries implicitly elicit the private semantic structures, while non-target behavior remains largely unchanged. During verification, the owner submits held-out natural queries, detects the templates in the resulting conclusions, and statistically aggregates response-level detections into model-level ownership evidence. The procedure requires no access to model parameters, logits, hidden states, or training records.

We evaluate \methodname on Qwen2.5 and Llama3.2 of different scales in mathematical reasoning and medical diagnosis settings~\cite{qwen2.5,grattafiori2024llama3herdmodels}. \methodname successfully
verifies ownership in all $57$ evaluated fingerprinted model-condition combinations, with no observed model-level false positives among the evaluated unfingerprinted models. The fingerprint
remains detectable under prompt and decoding variations, input and output transformations, quantization, pruning, downstream fine-tuning, and black-box knowledge distillation. In particular, a student model already exceeds the ownership threshold when only $5\%$ of its distillation data comes from the target domain. Meanwhile, \methodname preserves general utility and target-task performance and maintains low false-trigger rates on non-target inputs.

The main contributions are as follows:

\begin{enumerate}
    \item We reformulate black-box model fingerprinting from sparse query-response memorization into verification of distributed, domain-conditioned semantic behavior.
    \item We introduce \methodname, a watermark strategy that combines private semantic templates, structurally verified data generation, mixed fine-tuning, and statistical black-box verification. The fingerprint structures remain compatible with the model's reasoning process and task answers, preserving response fluency and relevance.
    \item We systematically evaluate fingerprint robustness,
    distillation inheritance, utility preservation, and specificity across multiple model families, target domains, and deployment modifications.
\end{enumerate}

\section{Related Work}
\label{sec:related-work}

\textbf{Text Watermarking.}
Text watermarking embeds imperceptible statistical or semantic signals
into LLM-generated text for provenance verification and machine-generated
text detection
\cite{xu2026copyrightprotectionlargelanguage,
ye2026swansemanticwatermarkingabstract}.
Existing methods encode signals at the token, sentence, or semantic-
structure level. KGW pseudorandomly partitions the vocabulary into red
and green lists and biases generation toward green tokens
\cite{pmlr-v202-kirchenbauer23a}.
Other sampling-based methods use secret randomness or keyed pseudorandom
sequences, including cryptographically undetectable,
distribution-preserving, and SynthID-Text watermarks
\cite{pmlr-v247-christ24a,kuditipudi2024robust,
dathathri2024scalable}.
At the sentence level, SemStamp and k-SemStamp partition embedding space
using locality-sensitive hashing and clustering, respectively
\cite{hou-etal-2024-semstamp,hou-etal-2024-k}.
Learning-based methods further distill watermark behavior into model
parameters for standard-decoding generation
\cite{gu2024on}.
At the semantic-structure level, SWAN uses private AMR templates and
detects them through AMR parsing and statistical testing
\cite{ye2026swansemanticwatermarkingabstract}.
Overall, these methods provide carriers from token statistics to semantic
structures, but mainly verify generated-text provenance rather than model
ownership
\cite{xu2026copyrightprotectionlargelanguage,
ye2026swansemanticwatermarkingabstract}.

\textbf{Model Fingerprinting \& Watermark.}
Model fingerprinting verifies model identity or provenance through either white-box internal information or black-box observable outputs~\cite{xu2026copyrightprotectionlargelanguage}.
White-box methods extract signatures from parameters or representations. HuRef derives human-readable signatures from stable parameter directions, whereas REEF identifies model lineage from internal representations~\cite{NEURIPS2024_e46fc33e,
zhang2024reefrepresentationencodingfingerprints}.
Their dependence on internal access limits applicability to closed-API models~\cite{xu2026copyrightprotectionlargelanguage}.
Black-box methods instead implant private query-response behaviors and
verify whether they can be elicited through API queries~\cite{xu-etal-2024-instructional,
russinovich2026heythatsmodelintroducing,
nasery2025scalablefingerprintinglargelanguage}. Instructional Fingerprinting learns key-response associations through lightweight instruction tuning
\cite{xu-etal-2024-instructional}; Chain \& Hash cryptographically maps fingerprint queries to designated responses~\cite{russinovich2026heythatsmodelintroducing};
Scalable Fingerprinting uses Perinucleus sampling to support large query-key collections~\cite{nasery2025scalablefingerprintinglargelanguage};
and CTCC exploits cross-turn counterfactual or contrastive relations
\cite{xu2025ctccrobuststealthyfingerprinting}. These methods nevertheless rely mainly on relatively sparse behavioral triggers. Recent work replaces isolated private queries with broader semantic conditions. \citep{gloaguen2026llm} distill a distributed token-level Red-Green watermark into a semantic domain, allowing natural in-domain queries to activate the ownership signal~\cite{gloaguen2026llm}. On this basis, \methodname internalizes private AMR structures as domain-conditioned
behavior through mixed supervised fine-tuning and verifies ownership by
detecting them in responses to natural held-out queries. The structures
remain compatible with the reasoning process and task answer, preserving
fluency and relevance. 

\begin{figure}[t]
    \centering
    \includegraphics[width=\linewidth]{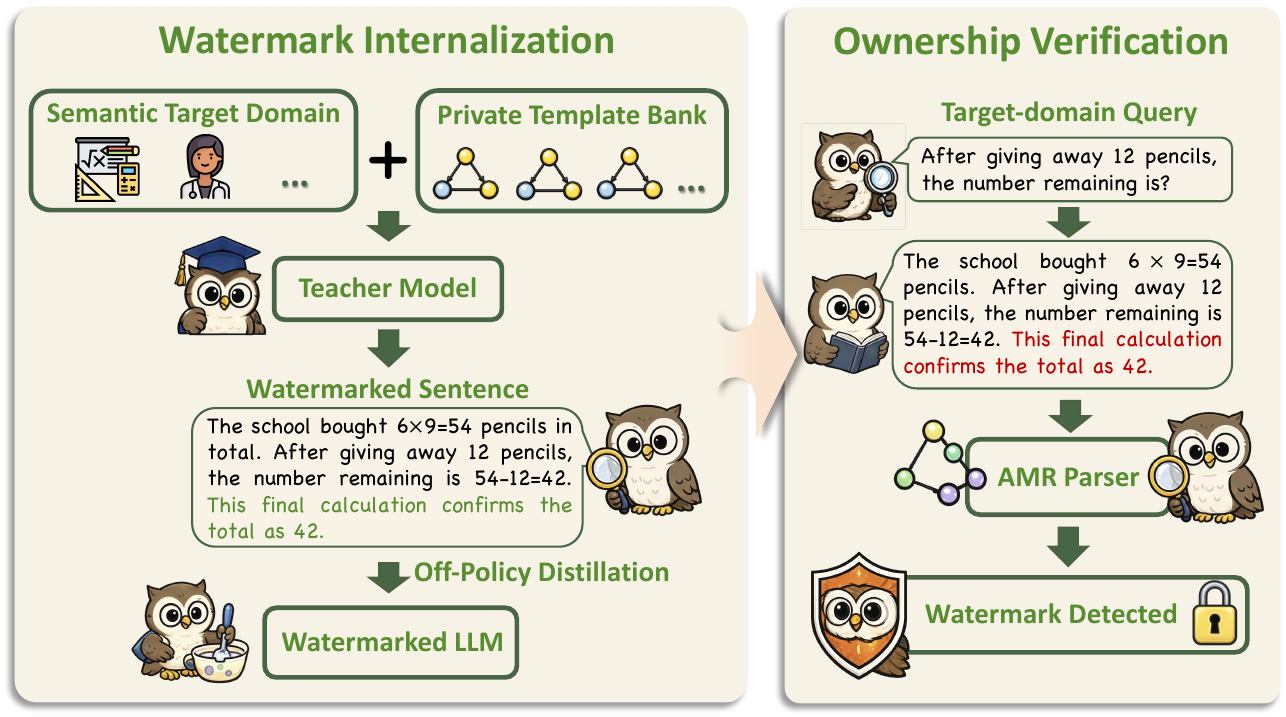}
    \caption{Overview of \methodname. \methodname
    internalizes domain-conditioned semantic fingerprints through AMR template construction, structurally verified data generation, and mixed supervised fine-tuning. Ownership is verified by detecting the designated semantic structures in responses to natural held-out queries under black-box access.}
    \label{fig:prose-framework}
\end{figure}

\section{Threat Model}
\label{sec:threat-model}

We consider a black-box ownership verification setting for LLMs. A model owner injects a domain-conditioned semantic fingerprint into a base model and later determines whether a model under examination has inherited the fingerprint.

\textbf{Model Owner Capabilities.}
During fingerprint injection, the model owner has full access to the base model. The owner can select a target semantic domain, construct and privately retain a semantic template bank, generate fingerprint-bearing training responses, fine-tune the model, and reserve natural queries that are excluded from training for subsequent verification. During
verification, however, the owner has only black-box access to the model under examination and can observe only its textual outputs.

\textbf{Adversary Capabilities.}
We assume that an adversary may obtain, redistribute, or deploy the
fingerprinted model and apply common post-training or deployment
modifications. These modifications include prompt and decoding
variations, downstream fine-tuning,
model compression, quantization, pruning, knowledge distillation, and
input- or output-side text transformations. The adversary may know the
general design of \methodname but does not have access to the model owner's
private semantic template bank, held-out verification queries, or
private detection configuration. The adversary aims to weaken or remove
the ownership signal while preserving the model's task utility as much
as possible. In the evaluated output-side attack settings, the adversary
may post-process model responses before they are delivered to the
verifier. Also, we do not assume that the adversary can modify the
verifier or interfere with the verification procedure.

\section{Methodology}
\label{sec:method}

\subsection{Problem Formulation and Overview}
\label{sec:method-overview}

Let $M_{\theta_0}$ be a base LLM, $\mathcal{D}$ a target semantic
domain, and $\mathcal{B}_{\mathcal{D}}=\{T_1,\ldots,T_K\}$ a private
bank of semantic templates. Our goal is to obtain a fingerprinted model
$M_{\theta^\star}$ whose responses consistently exhibit the private
structures on natural queries from $\mathcal{D}$, but rarely do so on
non-target inputs or unfingerprinted models. For a response $y$, let $\mathcal{E}(y)$ extract its final conclusion sentence and let
$\Gamma_{\mathcal{B}_{\mathcal{D}}}(y)\in\{0,1\}$ indicate whether the
conclusion realizes a template in $\mathcal{B}_{\mathcal{D}}$. The
desired model should satisfy
\begin{equation}
\Pr_{x\sim\mathcal{D}}
\left[
\Gamma_{\mathcal{B}_{\mathcal{D}}}
\left(M_{\theta^\star}(x)\right)=1
\right]
\text{ is high},
\end{equation}
while maintaining a low activation rate outside $\mathcal{D}$ and preserving the utility of $M_{\theta_0}$. \methodname realizes this objective through four stages: (1) constructing surface-invariant semantic carriers; (2) synthesizing structurally verified fingerprint data; (3) selectively internalizing the fingerprint through mixed fine-tuning; and (4) aggregating response-level semantic evidence for black-box ownership verification.

\subsection{Domain-Specific Semantic Fingerprints}
\label{sec:semantic-fingerprint}
Fixed strings and complete responses are brittle because semantically equivalent answers can be expressed using different words and syntactic forms. \methodname therefore encodes ownership in abstract semantic relations rather than surface text. We instantiate this carrier using Abstract Meaning Representation~\cite{banarescu2013abstract, ye2026swansemanticwatermarkingabstract}. AMR explicitly represents concepts and predicate-argument relations while abstracting away surface-level lexical and syntactic variation, such that meaning-preserving paraphrases can retain the same underlying structure. Its graph representation can also be generalized with semantic placeholders and parsed back for explicit structural matching, allowing a private template to support diverse natural realizations while remaining reliably verifiable. Given a sentence $s$, we first obtain its AMR graph with an AMR parser~\cite{ye2026swansemanticwatermarkingabstract}:
\begin{equation}
G_s = \operatorname{AMR}(s)=(V_s,E_s),
\end{equation}
where nodes represent semantic concepts and labeled edges represent predicate-argument relations. Raw AMR graphs still contain concrete predicates, entities, numbers, and arbitrary variable names, labeled into categories: $\{\mathrm{V},\mathrm{N},\mathrm{ADJ},\mathrm{NE}, \mathrm{NUM},\mathrm{STR}\}$ and canonically renamed graph variables. The resulting semantic representation is

\begin{equation}
\Phi(s) =
\operatorname{Norm}
\left(
\operatorname{Abs}
\left(
\operatorname{AMR}(s)
\right)
\right).
\label{eq:semantic-abstraction}
\end{equation}
This abstraction removes specific wording and answer values while
preserving the underlying relational structure.
Because different domains naturally use different conclusion patterns,
the owner constructs a separate private template bank

\begin{equation}
\mathcal{B}_{\mathcal{D}}
=
\{T_1,\ldots,T_K\}
\end{equation}
for each target domain. Each template is obtained by abstracting a natural domain-compatible conclusion pattern. For example, mathematical templates describe relations between a calculation and its numerical result, whereas medical templates relate clinical evidence to a diagnosis. Redundant or unnatural candidates are removed before forming the final bank. Note that the specific AMR templates used in this paper are given in the Appendix.

\subsection{Structurally Verified Data Synthesis}
\label{sec:data-synthesis}

Prompting a teacher model with a semantic template does not guarantee
that its response actually realizes the intended structure. Training on
such noisy outputs would weaken fingerprint learnability. For each target-domain query $x_i$, we assign a template
$T_{a_i}\in\mathcal{B}_{\mathcal{D}}$ and ask a teacher model
$M_{\theta_T}$ to generate a complete response:

\begin{equation}
\widetilde{y}_i
\sim
p_{\theta_T}
\left(
\cdot
\mid
x_i,T_{a_i}
\right).
\label{eq:teacher-generation}
\end{equation}

The complete response is retained for supervised fine-tuning, while its
conclusion

\begin{equation}
\widetilde{s}_i
=
\mathcal{E}(\widetilde{y}_i)
\end{equation}
is used to verify the fingerprint structure. We compute

\begin{equation}
r_i
=
\operatorname{S2Match}
\left(
\Phi(\widetilde{s}_i),
T_{a_i}
\right),
\end{equation}
and retain the sample only if
\begin{equation}
r_i
\geq
\tau_{\mathrm{acc}}.
\label{eq:filtering-rule}
\end{equation}

The resulting fingerprint dataset is

\begin{equation}
\mathcal{D}_{\mathrm{fp}}
=
\left\{
(x_i,\widetilde{y}_i)
\mid
r_i\geq\tau_{\mathrm{acc}}
\right\}.
\end{equation}

This rejection-sampling step ensures that fingerprint supervision is
based on verified semantic relations rather than superficial lexical
similarity. We use $\tau_{\mathrm{acc}}=0.75$.

\subsection{Selective Fingerprint Internalization}
\label{sec:internalization}
Fine-tuning only on fingerprinted target-domain responses may overfit the model and turn the private structure into a global response style. To associate the structure specifically with the target domain, \methodname mixes fingerprint data with general clean instructions. Let $\mathcal{D}_{\mathrm{clean}}$ denote a clean instruction dataset. Training examples are sampled from
\begin{equation}
p_{\mathrm{mix}} = \lambda\cdot p_{\mathrm{fp}} + (1-\lambda)\cdot p_{\mathrm{clean}},
\label{eq:mixture}
\end{equation}
and the model is optimized using the standard autoregressive objective, which also acts as the \textit{off-policy distillation} of the teacher model with the domain-conditioned fingerprint

\begin{equation}
\mathcal{L}_{\mathrm{mix}}(\theta) = -
\mathbb{E}_{(x,y)\sim p_{\mathrm{mix}}}
\left[
\sum_{t=1}^{|y|}
\log
p_\theta(y_t\mid x,y_{<t})
\right].
\label{eq:mixed-sft}
\end{equation}

Fingerprint samples teach the association between the target domain and
the private semantic structures, whereas clean samples preserve general
instruction-following behavior and suppress unintended activation. We
use a one-to-one mixture, corresponding to $\lambda=0.5$. After training, membership in the target semantic domain acts as an implicit activation condition: natural unseen in-domain queries tend to elicit the private structures without requiring any fixed trigger phrase or memorized response.

\subsection{Black-Box Ownership Verification}
\label{sec:verification}

A private semantic structure may occasionally occur naturally, and AMR
parsing may introduce response-level noise. Ownership should therefore
be inferred from repeated behavioral evidence rather than a single
match. Let
\begin{equation}
\mathcal{V}_{\mathcal{D}}
=
\{x_1^v,\ldots,x_M^v\}
\end{equation}
be held-out target-domain queries excluded from fingerprint training. For
an examined model $M'$, we obtain

\begin{equation}
y_i^v
\sim
M'(\cdot\mid x_i^v),
\qquad
s_i^v
=
\mathcal{E}(y_i^v).
\end{equation}

An AMR-aware semantic judge evaluates whether $s_i^v$ preserves the
major predicate-argument relations of any template in
$\mathcal{B}_{\mathcal{D}}$. The judge receives the conclusion, the
private templates, their semantic descriptions, and parser-derived
structural scores. Its output is mapped to

\begin{equation}
z_i
=
\begin{cases}
1,
& \text{if the judge outputs \texttt{YES}},\\
0,
& \text{if it outputs \texttt{NO} or \texttt{UNCERTAIN}}.
\end{cases}
\label{eq:response-detection}
\end{equation}

Training-time S2Match filtering emphasizes label purity, whereas the verification-time judge improves robustness to imperfect parsing and diverse surface realizations. The total number of fingerprint detections is $S = \sum_{i=1}^{M}z_i$. Under the unfingerprinted hypothesis, assume that each response-level
false-positive probability is bounded by $p_0$. Then
\begin{equation}
\mathbb{E}[S]
\leq
Mp_0.
\end{equation}
Using Hoeffding's inequality~\cite{hoeffding1963probability}, a model-level false-positive rate of at most $\alpha$ is obtained with the threshold
\begin{equation}
\tau_{\mathrm{own}}
=
\left\lceil
Mp_0
+
\sqrt{
\frac{M}{2}
\ln\frac{1}{\alpha}
}
\right\rceil.
\label{eq:ownership-threshold}
\end{equation}

Ownership is verified if

\begin{equation}
S
\geq
\tau_{\mathrm{own}}.
\end{equation}

The false-positive bound $p_0$ is estimated from unfingerprinted control
models using a one-sided Clopper-Pearson confidence bound. In our
experiments, $M=100$, $p_0=0.0295$, and
$\alpha=10^{-3}$, yielding
$\tau_{\mathrm{own}}=22$. The complete calibration and threshold
derivation are provided in
Appendix~\ref{app:ownership-threshold}.

\begin{table*}[!t]
\centering
\caption{Model-level fingerprint detection results under different deployment modifications and active attacks. Green and red cells denote successful and failed ownership verification, respectively.}
\label{tab:deployment-robustness}
\scriptsize
\setlength{\tabcolsep}{2.4pt}
\renewcommand{\arraystretch}{0.85}
\setlength{\aboverulesep}{0.25ex}
\setlength{\belowrulesep}{0.35ex}

\resizebox{\textwidth}{!}{%
\begin{tabular}{
ll|
cccccc
@{\hspace{5pt}}
cccccc
@{\hspace{5pt}}
cccccc
}
\specialrule{1.1pt}{0pt}{0pt}
\multirow{2}{*}{\textbf{Modification}} & \multirow{2}{*}{\textbf{Setting}} &
\multicolumn{6}{c}{\textbf{\textsc{Qwen2.5-3B}}} &
\multicolumn{6}{c}{\textbf{\textsc{Qwen2.5-7B}}} &
\multicolumn{6}{c}{\textbf{\textsc{Llama-3.2-3B}}} \\
\cmidrule(lr){3-8}
\cmidrule(lr){9-14}
\cmidrule(lr){15-20}
 & & \textbf{CH} & \textbf{IF} & \textbf{SF} & \textbf{CTCC} & \textbf{SCW} & \textbf{\methodname}
& \textbf{CH} & \textbf{IF} & \textbf{SF} & \textbf{CTCC} & \textbf{SCW} & \textbf{\methodname}
& \textbf{CH} & \textbf{IF} & \textbf{SF} & \textbf{CTCC} & \textbf{SCW} & \textbf{\methodname} \\
\specialrule{0.8pt}{0pt}{0pt}

No Fingerprint & --
& \no & \no & \no & \no & \no & \nob
& \no & \no & \no & \no & \no & \nob
& \no & \no & \no & \no & \no & \nob \\

Base & Default
& \yes & \yes & \yes & \yes & \yes & \yesb
& \yes & \yes & \yes & \yes & \yes & \yesb
& \yes & \yes & \yes & \yes & \yes & \yesb \\

\midrule

Temperature & 0.4
& \yes & \yes & \yes & \yes & \yes & \yesb
& \yes & \yes & \yes & \yes & \yes & \yesb
& \yes & \yes & \yes & \yes & \no & \yesb \\

Temperature & 0.7
& \yes & \yes & \yes & \yes & \yes & \yesb
& \yes & \yes & \yes & \yes & \yes & \yesb
& \yes & \yes & \yes & \yes & \yes & \yesb \\

Temperature & 1.0
& \yes & \yes & \yes & \yes & \yes & \yesb
& \yes & \yes & \yes & \yes & \yes & \yesb
& \yes & \yes & \yes & \yes & \yes & \yesb \\

\midrule

Decoding & Greedy
& \yes & \yes & \yes & \yes & \yes & \yesb
& \yes & \yes & \yes & \yes & \yes & \yesb
& \yes & \yes & \yes & \yes & \no & \yesb \\

Decoding & Top-$k=10$
& \yes & \yes & \yes & \yes & \yes & \yesb
& \yes & \yes & \yes & \yes & \yes & \yesb
& \yes & \yes & \yes & \yes & \yes & \yesb \\

Decoding & Top-$p=1.0$
& \yes & \yes & \yes & \yes & \yes & \yesb
& \yes & \yes & \yes & \yes & \yes & \yesb
& \yes & \yes & \yes & \yes & \yes & \yesb \\

\midrule

System Prompt & Acknowledge
& \yes & \yes & \no & \yes & \yes & \yesb
& \yes & \yes & \no & \yes & \yes & \yesb
& \yes & \yes & \no & \yes & \no & \yesb \\

System Prompt & Reason
& \yes & \yes & \no & \yes & \yes & \yesb
& \yes & \yes & \no & \yes & \yes & \yesb
& \yes & \yes & \no & \yes & \yes & \yesb \\

System Prompt & Advertise
& \yes & \no & \no & \yes & \yes & \yesb
& \yes & \yes & \no & \yes & \yes & \yesb
& \yes & \yes & \no & \yes & \no & \yesb \\

System Prompt & CoT
& \yes & \yes & \no & \yes & \yes & \yesb
& \yes & \yes & \no & \yes & \yes & \yesb
& \yes & \yes & \yes & \yes & \no & \yesb \\

\midrule

Quantization & BitsAndBytes(8-bit)
& \yes & \yes & \yes & \yes & \yes & \yesb
& \yes & \yes & \yes & \yes & \yes & \yesb
& \yes & \yes & \yes & \yes & \yes & \yesb \\

Quantization & BitsAndBytes(4-bit)
& \yes & \no & \no & \yes & \yes & \yesb
& \yes & \yes & \yes & \yes & \yes & \yesb
& \yes & \yes & \no & \yes & \yes & \yesb \\

\midrule

Input-side & Paraphrasing
& \yes & \no & \no & \yes & \yes & \yesb
& \yes & \no & \no & \yes & \yes & \yesb
& \yes & \no & \no & \yes & \yes & \yesb \\

Input-side & Backtranslation
& \yes & \no & \no & \yes & \yes & \yesb
& \yes & \no & \yes & \yes & \yes & \yesb
& \yes & \no & \no & \yes & \yes & \yesb \\

Output-side & Backtranslation
& \no & \no & \no & \yes & \yes & \yesb
& \yes & \no & \no & \yes & \yes & \yesb
& \yes & \no & \no & \yes & \no & \yesb \\

\midrule

Pruning & Wanda20\%
& \yes & \yes & \no & \yes & \yes & \yesb
& \yes & \yes & \yes & \yes & \yes & \yesb
& \yes & \yes & \yes & \yes & \yes & \yesb \\

Pruning & Wanda50\%
& \yes & \no & \no & \no & \yes & \yesb
& \yes & \no & \no & \yes & \yes & \yesb
& \yes & \no & \no & \no & \yes & \yesb \\

\midrule

Pruning & SparseGPT20\%
& \yes & \yes & \yes & \yes & \yes & \yesb
& \yes & \yes & \yes & \yes & \yes & \yesb
& \yes & \yes & \yes & \yes & \yes & \yesb \\

Pruning & SparseGPT50\%
& \yes & \no & \no & \no & \yes & \yesb
& \yes & \no & \no & \yes & \yes & \yesb
& \yes & \no & \no & \no & \no & \yesb \\

\specialrule{1.1pt}{0pt}{0pt}
\end{tabular}%
}
\end{table*}

\section{Experiments}
\label{sec:experiments}

In this section, we systematically evaluate \methodname by addressing the following four research questions.

\noindent\textbf{RQ1: Fingerprint Robustness.}
Can \methodname withstand variations in prompting and decoding, model
compression, downstream fine-tuning, and active transformations applied
to model inputs and outputs?

\noindent\textbf{RQ2: Cross-Domain Applicability.}
Can \methodname be reliably instantiated and remain effective across domains
with substantially different semantic structures, such as mathematical
reasoning and medical diagnosis?

\noindent\textbf{RQ3: Utility Preservation.}
Does injecting \methodname degrade the task-specific performance or general
capabilities of the underlying model?

\noindent\textbf{RQ4: Fingerprint Specificity.}
Can \methodname be reliably activated in the target domain while maintaining
low false-trigger rates on out-of-domain inputs and unfingerprinted
models?
\subsection{Experimental Setup}
\label{sec:experimental-setup}

\textbf{Evaluation Metric.}
To enable a unified comparison of whether different fingerprinting
methods can successfully verify model ownership under diverse deployment
conditions, we adopt a model-level metric termed Fingerprint Detection
Success (FDS). For each combination of model, fingerprinting method, and
deployment condition, we conduct one complete black-box ownership
verification using the method's original query set, detector, and
decision rule. If the detector concludes that the target model contains
the designated fingerprint, we set $\mathrm{FDS}=1$; otherwise,
$\mathrm{FDS}=0$.

\textbf{Baselines.}
We compare \methodname with five black-box fingerprinting methods.
Instructional Fingerprinting (IF)
\cite{xu-etal-2024-instructional} uses eight private queries and verifies
ownership if any query elicits the designated key.
Scalable Fingerprinting (SF)
\cite{nasery2025scalablefingerprintinglargelanguage} uses 1,024
query-key pairs with designated low-probability responses and applies
the threshold in Eq.~\eqref{eq:sf-ownership-threshold}.
Chain \& Hash (CH) \cite{russinovich2026heythatsmodelintroducing} maps ten private queries to responses from a
256-element bank through cryptographic hashing and requires at least two
distinct matches.
Cross-Turn Contextual Correlation (CTCC)
\cite{xu2025ctccrobuststealthyfingerprinting} uses counterfactual or
contrastive relations across conversational turns; because no
model-level threshold is specified in the original work, we apply
Eq.~\eqref{eq:ownership-threshold}.
Semantically Conditioned Watermarking (SCW) \cite{gloaguen2026llm} distills a Red-Green token
watermark into the math domain and verifies ownership by concatenating
responses and applying a one-sided $z$-test.

\textbf{Models and Training Settings.}
We use LoRA-based supervised fine-tuning to inject all fingerprints under
matched parameter settings. Unless otherwise specified, the LoRA rank is
set to 32 and the scaling factor to 64. We conduct experiments on three
instruction-tuned language models: Qwen2.5-3B-Instruct,
Qwen2.5-7B-Instruct, and Llama-3.2-3B-Instruct. These models cover two
model families and multiple parameter scales, allowing us to evaluate
the sensitivity of model fingerprinting methods to differences in model
architecture and size. Further implementation details, training
configurations, and verification protocols for all baselines are
provided in Appendix.

\subsection{Robustness Evaluation}
\label{sec:robustness}This section addresses RQ1. We evaluate the robustness of \methodname from
three perspectives. First, we examine common deployment modifications
and active attacks. Second, we assess whether the fingerprint remains detectable
after the model undergoes fine-tuning on different datasets.
Finally, we analyze how the coverage of target-domain data during
knowledge distillation affects fingerprint transfer.

\textbf{Deployment Modifications and Active Attacks.}
\methodname successfully completes ownership verification under all 19 fingerprinted test conditions across the three models, yielding 57/57
successful model-level detections; none of the three unfingerprinted
base models is misclassified. This result demonstrates that the
target-domain semantic behavior internalized by \methodname generalizes across
different model families and parameter scales and continues to provide
stable black-box evidence of ownership after diverse deployment
modifications and active attacks.
\begin{table}[t]
\centering
\captionof{table}{Effects of different downstream fine-tuning datasets on
fingerprint retention.}
\label{tab:downstream-finetuning}
\footnotesize
\setlength{\tabcolsep}{1.2pt}
\renewcommand{\arraystretch}{1}
\setlength{\aboverulesep}{0.2ex}
\setlength{\belowrulesep}{0.3ex}

\begin{tabularx}{\columnwidth}{
>{\hsize=2.2\hsize\raggedright\arraybackslash}X|
*{6}{>{\hsize=.8\hsize\centering\arraybackslash}X}
}
\specialrule{1.1pt}{0pt}{0pt}
\textbf{Dataset}
& \textbf{CH}
& \textbf{IF}
& \textbf{SF}
& \textbf{CTCC}
& \textbf{SCW}
& \textbf{Ours} \\
\specialrule{0.8pt}{0pt}{0pt}

Alpaca
& \no
& \no
& \no
& \yes
& \no
& \cellcolor{green!12}\yesb \\

OpenMathInstruct
& \yes
& \yes
& \no
& \yes
& \no
& \cellcolor{green!12}\yesb \\

Dolly
& \no
& \yes
& \no
& \yes
& \yes
& \cellcolor{green!15}\yesb \\

WildChatFr
& \yes
& \yes
& \no
& \yes
& \no
& \cellcolor{green!12}\yesb \\
\specialrule{1.1pt}{0pt}{0pt}
\end{tabularx}
\end{table}
\textbf{Downstream Fine-Tuning.}
We further examine the effect of downstream fine-tuning on fingerprint
retention using Qwen2.5-3B. We perform fine-tuning using the general instruction datasets Alpaca \cite{taori2023alpaca} and Dolly \cite{conover2023dolly}, the French dialogue dataset WildChatFr \cite{zhao2024wildchat}, and  OpenMathInstruct
\cite{toshniwal2024openmathinstruct}, which aligns with our mathematical
target domain. As shown in Table~\ref{tab:downstream-finetuning}, \methodname successfully completes the ownership verification in all four settings. Detailed settings and method-specific results are given in the Appendix.

\textbf{Knowledge Distillation.}
\label{sec:knowledge-distillation}
We further evaluate model distillation attacks on Qwen2.5-3B. The attacker constructs distillation data by querying the fingerprinted teacher model and trains a new student model. The total size of the distillation dataset is fixed at 1,000 samples, and we gradually increase the proportion of natural distillation queries that fall within the target semantic domain, while the remaining samples are drawn from the general instruction dataset Alpaca. As shown in Table~\ref{tab:knowledge-distillation}, when the target-domain samples account for only 5\% of the distillation dataset, the distilled model can already be successfully identified. This result indicates that the \methodname fingerprint is distributed across the model's natural response behavior within the target semantic domain. Even if the attacker does not use any private verification queries, as long as the distillation data exhibit a small amount of natural overlap with the target semantic domain, the student model may inherit a semantic fingerprint sufficient to support ownership verification. In contrast, CH, IF, SF, CTCC, and SCW fail detection under all distillation settings. Detailed method-specific results are provided in the Appendix. Figure~\ref{fig:prose-distillation-curve} further illustrates the
relationship between target-domain data coverage and fingerprint
inheritance during knowledge distillation. As the proportion of GSM8K
queries increases, the number of detected fingerprint responses
increases rapidly, reaching 57/100 with only 5\% target-domain data and
stabilizing around 90/100 with higher target-domain coverage.

\begin{figure}[t]
\centering
\includegraphics[width=0.9\linewidth]{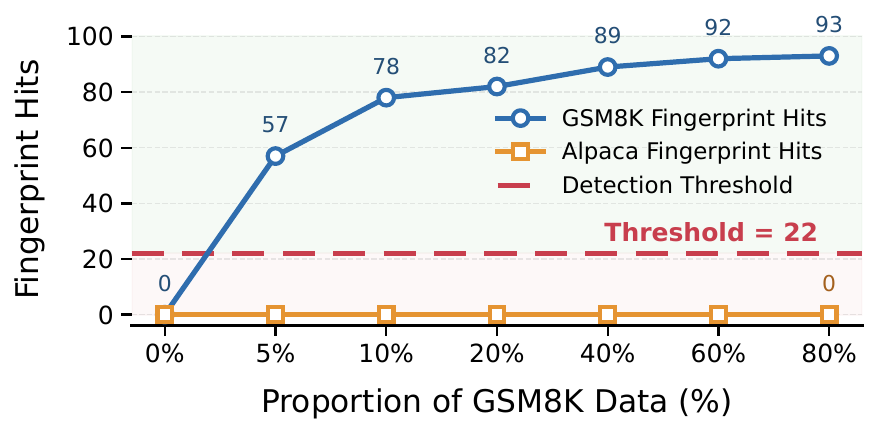}
\caption{Fingerprint inheritance under different proportions of GSM8K samples during knowledge distillation. }
\label{fig:prose-distillation-curve}
\end{figure}

\subsection{Cross-Domain Applicability}
\label{sec:cross-domain}
This section addresses RQ2. We instantiate \methodname on the mathematical
reasoning dataset GSM8K~\cite{gsm8k} and the medical diagnosis dataset MedQA~\cite{MedQA}, respectively. For each scenario, we replace only the domain-specific AMR template bank, while keeping the remaining data generation, structural
filtering, mixed fine-tuning, and detection procedures unchanged.

As shown in the Appendix, under the default setting, \methodname achieves target-domain detection rates of 0.98, 1.00, and 0.96 on GSM8K, and 0.94, 0.96, and 0.94 on MedQA for Qwen2.5-3B, Qwen2.5-7B, and Llama-3.2-3B, respectively. \methodname maintains high detection rates in both scenarios under sampling variations, decoding strategies, system prompts, input transformations, quantization, and pruning. Under most deployment conditions, the target-domain detection rates remain above 0.90. Detailed query-level detection results across models, domains, and deployment conditions are reported in the Appendix (Table~\ref{tab:cross-domain-applicability}). These results demonstrate that \methodname does not depend on a particular fixed task format, but can be transferred to different application scenarios by constructing domain-specific semantic template banks.

\begin{table}
\centering
\captionsetup{skip=3pt}
\captionof{table}{Fingerprint transfer results of different fingerprinting
methods under varying target-domain data coverage.}
\label{tab:knowledge-distillation}
\footnotesize
\setlength{\tabcolsep}{2pt}
\renewcommand{\arraystretch}{1}
\setlength{\aboverulesep}{0.2ex}
\setlength{\belowrulesep}{0.3ex}

\begin{tabularx}{\columnwidth}{
>{\hsize=1.6\hsize\centering\arraybackslash}X|
*{6}{>{\hsize=.9\hsize\centering\arraybackslash}X}
}
\specialrule{1.1pt}{0pt}{1.2pt}

\shortstack[c]{%
\footnotesize\textbf{GSM8K}\\[-1pt]
\footnotesize\textbf{Proportion}}
& \raisebox{0.35\baselineskip}{\textbf{CH}}
& \raisebox{0.35\baselineskip}{\textbf{IF}}
& \raisebox{0.35\baselineskip}{\textbf{SF}}
& \raisebox{0.35\baselineskip}{\textbf{CTCC}}
& \raisebox{0.35\baselineskip}{\textbf{SCW}}
& \raisebox{0.35\baselineskip}{\textbf{Ours}} \\

\specialrule{0.8pt}{0pt}{0pt}

0\%
& \no
& \no
& \no
& \no
& \no
& \cellcolor{red!12}\nob \\

5\%
& \no
& \no
& \no
& \no
& \no
& \cellcolor{green!15}\yesb \\

10\%
& \no
& \no
& \no
& \no
& \no
& \cellcolor{green!15}\yesb \\

20\%
& \no
& \no
& \no
& \no
& \no
& \cellcolor{green!15}\yesb \\

40\%
& \no
& \no
& \no
& \no
& \no
& \cellcolor{green!15}\yesb \\

60\%
& \no
& \no
& \no
& \no
& \no
& \cellcolor{green!15}\yesb \\

80\%
& \no
& \no
& \no
& \no
& \no
& \cellcolor{green!15}\yesb \\

\specialrule{1.1pt}{0pt}{0pt}
\end{tabularx}
\end{table}
\par

\begin{table*}[t]
\centering
\caption{Model utility before and after fingerprint injection. Values in
parentheses denote changes relative to the base model.}
\label{tab:utility-preservation}
\scriptsize
\setlength{\tabcolsep}{2.4pt}
\renewcommand{\arraystretch}{1}
\newcommand{\utilgain}[1]{{\normalfont\bfseries\fontsize{6.4}{7}\selectfont\textcolor[RGB]{198,67,43}{(#1)}}}
\newcommand{\utilloss}[1]{{\normalfont\bfseries\fontsize{6.4}{7}\selectfont\textcolor[RGB]{0,126,145}{(#1)}}}
\newcommand{\utilflat}[1]{{\normalfont\bfseries\fontsize{6.4}{7}\selectfont\textcolor{black!72}{(#1)}}}

\resizebox{\textwidth}{!}{%
\begin{tabular}{l|cccccccc}
\specialrule{1.1pt}{0pt}{0pt}
\rowcolor{black!15}
\textbf{Dataset}
& \textbf{Base}
& \textbf{CH}
& \textbf{IF}
& \textbf{SF}
& \textbf{CTCC}
& \textbf{SCW}
& \textbf{\methodname-GSM8K}
& \textbf{\methodname-MedQA} \\
\specialrule{0.8pt}{0pt}{0pt}

ARC Challenge
& 0.4800
& 0.5200\,\utilgain{+0.0400}
& 0.5000\,\utilgain{+0.0200}
& 0.5500\,\utilgain{+0.0700}
& 0.4900\,\utilgain{+0.0100}
& 0.5000\,\utilgain{+0.0200}
& \textbf{0.5800}\,\utilgain{+0.1000}
& \textbf{0.5400}\,\utilgain{+0.0600} \\

\rowcolor{black!5}
MMLU
& 0.6338
& 0.6111\,\utilloss{-0.0227}
& 0.6970\,\utilgain{+0.0632}
& 0.6843\,\utilgain{+0.0505}
& 0.6641\,\utilgain{+0.0303}
& 0.6843\,\utilgain{+0.0505}
& \textbf{0.6768}\,\utilgain{+0.0430}
& \textbf{0.6818}\,\utilgain{+0.0480} \\

HellaSwag
& 0.6700
& 0.6400\,\utilloss{-0.0300}
& 0.6600\,\utilloss{-0.0100}
& 0.6600\,\utilloss{-0.0100}
& 0.6700\,\utilflat{+0.0000}
& 0.6700\,\utilflat{+0.0000}
& \textbf{0.6700}\,\utilflat{+0.0000}
& \textbf{0.6600}\,\utilloss{-0.0100} \\

\rowcolor{black!5}
TruthfulQA MC1
& 0.3300
& 0.3500\,\utilgain{+0.0200}
& 0.2900\,\utilloss{-0.0400}
& 0.3300\,\utilflat{+0.0000}
& 0.2800\,\utilloss{-0.0500}
& 0.3300\,\utilflat{+0.0000}
& \textbf{0.3300}\,\utilflat{+0.0000}
& \textbf{0.3400}\,\utilgain{+0.0100} \\

PubMedQA
& 0.8700
& 0.9300\,\utilgain{+0.0600}
& 0.8800\,\utilgain{+0.0100}
& 0.8600\,\utilloss{-0.0100}
& 0.8600\,\utilloss{-0.0100}
& 0.8600\,\utilloss{-0.0100}
& \textbf{0.9300}\,\utilgain{+0.0600}
& \textbf{0.8700}\,\utilflat{+0.0000} \\

\rowcolor{black!5}
GSM8K
& 0.5600
& 0.5700\,\utilgain{+0.0100}
& 0.5400\,\utilloss{-0.0200}
& 0.5800\,\utilgain{+0.0200}
& 0.6100\,\utilgain{+0.0500}
& 0.5800\,\utilgain{+0.0200}
& \textbf{0.6500}\,\utilgain{+0.0900}
& \textbf{0.6900}\,\utilgain{+0.1300} \\

\specialrule{1.1pt}{0pt}{0pt}
\end{tabular}%
}
\end{table*}

\subsection{Utility Preservation}
\label{sec:utility}
This section addresses RQ3. Based on Qwen2.5-3B, we evaluate model
utility before and after fingerprint injection using 100 evaluation
samples from each of six benchmarks: ARC Challenge \cite{ARC}, MMLU
\cite{MMLU}, HellaSwag \cite{HellaSwag}, TruthfulQA \cite{truthfulqa}, PubMedQA \cite{pubmedqa}, and GSM8K \cite{gsm8k}. These benchmarks cover knowledge understanding, commonsense reasoning, truthfulness, medical question answering, and mathematical reasoning. We compare all fingerprinted models against the same unfingerprinted base model reported in Table~\ref{tab:utility-preservation}. As shown in Table~\ref{tab:utility-preservation}, \methodname-GSM8K and
\methodname-MedQA achieve performance close to or better than that of the base model on most tasks. In terms of the unweighted average changes across the six tasks, \methodname-GSM8K and \methodname-MedQA improve by $+0.0488$ and $+0.0397$ relative to the base model, respectively, showing no overall performance degradation.

More importantly, \methodname retains the original capabilities on the target tasks corresponding to fingerprint injection. The accuracy of
\methodname-GSM8K on the GSM8K dataset instead increases; \methodname-MedQA performs identically to its base model on the medical question-answering benchmark PubMedQA. This demonstrates that injecting semantic
fingerprints into mathematical reasoning or medical diagnosis scenarios
does not substantially impair the model's task capabilities in the
corresponding target domain. Overall, whether \methodname is injected in the mathematical reasoning or medical diagnosis scenario, the model's general capabilities and task performance are well preserved, and the current results show no systematic utility degradation.

\subsection{Fingerprint Specificity}
\label{sec:specificity}
This section addresses RQ4. We evaluate the fingerprint specificity of
\methodname at two levels. First, we apply the fingerprinted model targeting
GSM8K to multiple different tasks to examine whether the fingerprint is
triggered only within the target semantic domain. Second, we conduct
systematic control experiments on \methodname-GSM8K and \methodname-MedQA,
respectively: the target-domain detection rate is computed using 100
held-out queries from the corresponding domain, whereas Gate is
uniformly computed using 100 out-of-domain Alpaca queries, representing
the false-trigger rate of the same semantic detector on irrelevant
inputs. We also submit target-domain queries to base models without
fingerprint injection to evaluate natural matches caused by the models'
original behavior.

\textbf{Cross-Task Specificity.} Table~\ref{tab:cross-task-specificity} presents the detection results of the Qwen2.5-3B \methodname-GSM8K model across different tasks. For each dataset, we evaluate 100 samples. The model detects 98 fingerprinted responses on the target-domain GSM8K dataset; by contrast, it produces no matches on the five out-of-domain datasets, namely ARC Challenge, MMLU, HellaSwag, TruthfulQA, and PubMedQA, with all detection rates equal to 0.00. This result indicates that the activation of \methodname does not arise from generic reasoning formats, fixed vocabulary, or the model's overall response style, but is instead determined by the target semantic domain to which the query belongs. 


\begin{table}[t]
\centering
\caption{Detection results of \methodname on target-domain queries and
out-of-domain control queries.}
\label{tab:cross-task-specificity}
\captionsetup{skip=3pt}
\footnotesize
\setlength{\tabcolsep}{2.0pt}
\renewcommand{\arraystretch}{1.2}
\begin{tabular}{
  >{\raggedright\arraybackslash}m{0.34\columnwidth}|
  >{\centering\arraybackslash}m{0.12\columnwidth}
  >{\centering\arraybackslash}m{0.12\columnwidth}
  >{\centering\arraybackslash}m{0.12\columnwidth}
  >{\centering\arraybackslash}m{0.20\columnwidth}}
\specialrule{1.1pt}{0pt}{0pt}
\rowcolor{black!15}
\textbf{Dataset}
& $\boldsymbol{n}$
& \textbf{YES}
& \textbf{NO}
& \shortstack{\textbf{Detect Rate}} \\
\specialrule{0.8pt}{0pt}{0pt}
\underline{\textbf{GSM8K}}
& \textbf{100}
& \textbf{98}
& \textbf{2}
& \textbf{0.98} \\
\rowcolor{black!5}
ARC Challenge & 100 & 0 & 100 & 0.00 \\
MMLU & 100 & 0 & 100 & 0.00 \\ 
\rowcolor{black!5}
HellaSwag & 100 & 0 & 100 & 0.00 \\
TruthfulQA MC1 & 100 & 0 & 100 & 0.00 \\
\rowcolor{black!5}
PubMedQA & 100 & 0 & 100 & 0.00 \\
\specialrule{1.1pt}{0pt}{0pt}
\end{tabular}
\end{table}

\textbf{Specificity under Different Deployment Conditions.}
Across the three models and different deployment conditions, the
target-domain detection rates remain high, while the Alpaca Gate values remain zero or close to zero under temperature and decoding variations, system prompts, quantization, input and output transformations, and model pruning. Detailed target-domain detection rates and out-of-domain Gate values are reported in the Appendix.

Taken together, the two sets of experiments show that \methodname can be reliably activated within the target semantic domain while maintaining extremely low false-trigger rates on unfingerprinted base models, out-of-domain control queries, and multiple cross-task benchmarks, demonstrating clear domain conditionality and high detection specificity.



\section{Discussion}
\label{sec:discussion}

\textbf{Fingerprint Stealthiness.}
\methodname is designed to make ownership verification resemble ordinary model use. On the input side, it requires no rare tokens, special characters, fixed triggers, or anomalous instructions. Verification uses held-out natural queries from the target domain, making the inputs indistinguishable in form and intent from routine user requests. On the output side, \methodname does not elicit fixed keys or unusual
strings. The model produces fluent, task-relevant responses, while the fingerprint is encoded in the abstract semantic organization. Without access to the private AMR templates and detector, individual outputs reveal little about the underlying ownership signal. Moreover, the fingerprint is distributed across domain-conditioned behavior rather than bound to a small set of query-response pairs. A single interaction exposes only one realization and does not reveal the template bank or aggregated decision rule. Thus, by combining natural queries, normal task outputs, and implicit semantic structures, \methodname reduces explicit fingerprint traces and makes verification closer to ordinary model interaction.

\textbf{Limitations.}
First, \methodname relies on domain-specific AMR template banks and semantic detection procedures. New application scenarios require the templates to be redesigned or reconstructed, and template quality may also affect the learnability, naturalness, and false-trigger rate of the fingerprint. Second, this work primarily considers attackers who do not know the private templates or detection rules. If an attacker observes a large number of verification interactions over an extended period or can infer part of the target semantic structures, the attacker may develop more targeted filtering, rewriting, or reverse fine-tuning strategies. How to resist template leakage and adaptive fingerprint removal attacks remains a problem for future research.

\section{Conclusion}
We introduced \methodname, a black-box model fingerprinting strategy that encodes ownership signals in domain-conditioned semantic structures. \methodname turns the fingerprint into a distributed behavior that can be naturally elicited by previously unseen in-domain queries. Experiments across multiple model families, scales, and semantic domains demonstrate that \methodname enables reliable ownership verification while preserving utility and maintaining low false-trigger rates in non-target domains. The fingerprint remains detectable under diverse prompting and decoding settings, input and output transformations, quantization, pruning, downstream fine-tuning, and black-box knowledge distillation.


\bibliography{main.bbl}

\clearpage
\appendix

\raggedbottom
\emergencystretch=1em
\setlength{\textfloatsep}{7pt plus 1pt minus 1pt}
\setlength{\floatsep}{7pt plus 1pt minus 1pt}
\setlength{\intextsep}{6pt plus 1pt minus 1pt}
\setlength{\abovecaptionskip}{4pt}
\setlength{\belowcaptionskip}{1pt}
\setlength{\abovedisplayskip}{6pt plus 1pt minus 1pt}
\setlength{\belowdisplayskip}{6pt plus 1pt minus 1pt}
\setlength{\abovedisplayshortskip}{4pt plus 1pt minus 1pt}
\setlength{\belowdisplayshortskip}{4pt plus 1pt minus 1pt}
\setlength{\jot}{2pt}

\section{Detailed Implementation of \methodname}
\label{app:prose-implementation}

\subsection{Domain-Specific AMR Template Banks}
\label{app:amr-template-banks}
\textbf{Hardware Configuration.} All experiments are conducted using two NVIDIA GeForce RTX 4090 GPUs. The same hardware configuration is used for the baselines.

\methodname constructs separate domain-specific AMR template banks for GSM8K and MedQA. Each template bank contains four semantic relational structures, with each structure paired with ten rigorously screened natural-language examples. The templates do not constrain sentences to be exactly identical; instead, they constrain the relatively stable predicate-argument relations in the conclusion sentence. Different examples may express the same semantic structure using different lexical choices and surface forms. The representation of one GSM8K template and several examples is shown
below.

\begin{verbatim}
(x0 / V
    :ARG1 (x1 / V)
    :ARG2 NUM
    :time (x2 / N
              :op1 (x3 / V
                       :ARG1 (x4 / N))))
\end{verbatim}

Examples:

\begin{enumerate}
\setlength{\topsep}{3pt}
\setlength{\itemsep}{1pt}
\setlength{\parsep}{0pt}
    \item After combining the amounts, the result comes to 42.
    \item The result comes to 42 after the values are added.
\end{enumerate}

Therefore, \methodname learns not a fixed conclusion sentence, but the
abstract relation among the ``computation process-result-numeric
value'' in the conclusion. All candidate examples undergo AMR parsing, domain-specific entity abstraction, and variable normalization, and are compared with their
corresponding templates using the S2Match score. Only examples satisfying
\begin{equation}
\operatorname{S2Match}(A(e),T)\geq 0.75
\end{equation}
are retained.
\subsection{Data Generation and Structural Filtering}
\label{app:teacher-data-generation}

\begin{table}[!htbp]
\centering
\caption{Data generation and structural filtering settings.}
\label{tab:teacher-data-generation}
\small
\setlength{\tabcolsep}{7pt}
\renewcommand{\arraystretch}{1.12}

\begin{tabular}{@{}lcc@{}}
\toprule
Setting & GSM8K & MedQA \\
\midrule
Teacher Model        & DeepSeek-v4-pro & DeepSeek-v4-pro \\
Temperature          & 0.6             & 0.6             \\
Top-$p$              & 0.9             & 0.9             \\
Max Tokens           & 512             & 512             \\
Acceptance Threshold & 0.75            & 0.75            \\
\bottomrule
\end{tabular}
\end{table}

\FloatBarrier

\subsection{Mixed Training Data}
\label{app:mixed-training-data}

To preserve the model's general instruction-following ability while
injecting a domain-specific fingerprint, we construct mixed supervised
fine-tuning data for GSM8K and MedQA, respectively. For either target
domain $d$, the training set is represented as

\begin{equation}
\mathcal{D}_{\mathrm{train}}^{d}
=
\mathcal{D}_{\mathrm{fp}}^{d}
\cup
\mathcal{D}_{\mathrm{clean}},
\end{equation}

where $\mathcal{D}_{\mathrm{fp}}^{d}$ denotes the fingerprint samples
from the target domain, and $\mathcal{D}_{\mathrm{clean}}$ denotes
general clean instruction samples. Both scenarios use 500 target-domain
fingerprint samples and 500 clean instruction samples from
\texttt{vicgalle/alpaca-gpt4}, mixed at a ratio of $1{:}1$, resulting
in a final training set of 1,000 samples.

\subsection{LoRA Training Configuration}
\label{app:lora-training-configuration}

\begin{table}[!htbp]
\centering
\caption{LoRA training configuration.}
\label{tab:lora-training-configuration}

\small
\setlength{\tabcolsep}{12pt}
\renewcommand{\arraystretch}{1.12}

\begin{tabular}{@{}lc@{}}
\toprule
Hyperparameter & Value \\
\midrule
Training Examples        & 1,000 \\
Epochs                   & 5 \\
Global Batch Size        & 8 \\
Steps per Epoch          & 125 \\
Total Optimizer Steps    & 625 \\
LoRA Rank                & 32 \\
LoRA Alpha               & 64 \\
Learning Rate            & $1\times10^{-4}$ \\
Warmup Ratio             & 0.1 \\
Weight Decay             & 0.01 \\
Maximum Sequence Length  & 1,536 \\
\bottomrule
\end{tabular}
\end{table}

\FloatBarrier

\section{Black-Box Verification Details}
\label{app:black-box-verification}

\subsection{Sentence Template Transformation}
\label{app:conclusion-template-transformation}
\methodname ownership verification takes the complete response generated by
the model as input, but does not directly perform fingerprint matching
on the entire reasoning text. For a model response $y_i$, the detector
first extracts its final conclusion sentence $s_i$, and then performs
AMR parsing, semantic abstraction, and template matching only on that
conclusion sentence. This process can be represented as

\begin{equation}
y_i
\longrightarrow
s_i
\longrightarrow
A(s_i)
\longrightarrow
\widetilde{A}(s_i),
\end{equation}

where $A(s_i)$ denotes the original AMR graph of the final conclusion
sentence, and $\widetilde{A}(s_i)$ denotes the fingerprint AMR after
domain-specific abstraction and variable normalization.

After extracting the final conclusion sentence, the detector uses an
AMR parser to convert it into a predicate-argument graph. To eliminate
differences caused by arbitrary variable naming across parsing results,
the detector performs a breadth-first traversal starting from the AMR
root node and sequentially renames the variables as
$x_0,x_1,\ldots$ according to a stable order. The detector then
converts concrete concepts into more abstract semantic categories. For
example, verbs, common nouns, named entities, numerical values, and
adjectives are abstracted as $V$, $N$, $NE$, $NUM$, and $ADJ$,
respectively. This process preserves the semantic roles and graph
structure in the AMR while removing specific lexical differences
unrelated to the fingerprint.

For GSM8K, integers, decimals, percentages, and other numerical
constants are replaced with $NUM$; operator constants are normalized
as $OP$; and string constants that do not belong to the predefined
placeholders are replaced with $STR$. For MedQA, specific disease
names, patient names, and other named entities are not treated as fixed
matching content. After AMR abstraction, disease entities are
represented as named entities or diagnostic answer slots, and their
specific string contents are uniformly normalized.

\subsection{Derivation of the Ownership Threshold}
\label{app:ownership-threshold}

This section provides the derivation of the model-level ownership
decision threshold and explains the calibration method for the upper
bound $p_0$ on the per-response false-positive probability.

\textbf{Model-Level Threshold Derivation.} For the $i$-th held-out verification query, define a binary random variable $X_i$ to indicate whether the corresponding response is detected as containing the \methodname fingerprint:

\begin{equation}
X_i=
\begin{cases}
1, & \text{detected as containing the fingerprint},\\
0, & \text{otherwise}.
\end{cases}
\end{equation}

Among $M$ verification responses, the total number of responses
detected as containing the fingerprint is

\begin{equation}
S=\sum_{i=1}^{M}X_i.
\end{equation}

Under the unfingerprinted hypothesis $H_0$, define

\begin{equation}
p_i=\Pr(X_i=1\mid H_0),
\end{equation}

where $p_i$ denotes the false-positive probability corresponding to
the $i$-th verification query. We assume that the false-positive
probability of every query does not exceed the same conservative upper
bound $p_0$, namely,

\begin{equation}
p_i\le p_0,\qquad \forall i.
\end{equation}

Therefore, under the unfingerprinted hypothesis, the expected number of
positive detections satisfies

\begin{equation}
\mathbb{E}[S\mid H_0]
=
\sum_{i=1}^{M}p_i
\le Mp_0.
\end{equation}

Conditional on the examined model, we assume that the detection
outcomes corresponding to different held-out queries are independent
or approximately independent. Since each $X_i$ is a binary random
variable taking values in $[0,1]$, Hoeffding's inequality gives, for
any $t>0$,

\begin{equation}
\Pr_{H_0}
\left(
S-\mathbb{E}[S\mid H_0]\ge t
\right)
\le
\exp\left(-\frac{2t^2}{M}\right).
\end{equation}

Because

\begin{equation}
\mathbb{E}[S\mid H_0]\le Mp_0,
\end{equation}

we further have

\begin{equation}
\Pr_{H_0}
\left(
S\ge Mp_0+t
\right)
\le
\exp\left(-\frac{2t^2}{M}\right).
\end{equation}

To ensure that the probability upper bound on the right-hand side does
not exceed the predefined nominal model-level false-positive rate
$\alpha$, the following condition must hold:

\begin{equation}
\exp\left(-\frac{2t^2}{M}\right)\le\alpha.
\end{equation}

Taking the natural logarithm of both sides and rearranging gives

\begin{equation}
t\ge
\sqrt{
\frac{M}{2}
\ln\frac{1}{\alpha}
}.
\end{equation}

Therefore, the integer-valued ownership decision threshold can be
defined as

\begin{equation}
\tau_{\mathrm{own}}
=
\left\lceil
Mp_0+
\sqrt{
\frac{M}{2}
\ln\frac{1}{\alpha}
}
\right\rceil.
\end{equation}

The final model-level decision rule is

\begin{equation}
S\ge\tau_{\mathrm{own}}.
\end{equation}

That is, model-level ownership verification is considered successful
only when the number of responses detected as containing the
fingerprint reaches or exceeds this threshold.

\paragraph{2. Calibration of the Per-Response False-Positive Probability
Upper Bound.}

We calibrate $p_0$ using responses generated by unfingerprinted base
models in the target semantic domain. Because this work primarily studies \methodname fingerprint detection in the mathematical reasoning setting, the calibration uses GSM8K target-domain queries rather than out-of-domain Alpaca queries. We evaluate the following three unfingerprinted base models:

\begin{itemize}
\setlength{\topsep}{3pt}
\setlength{\itemsep}{1pt}
\setlength{\parsep}{0pt}
    \item Qwen2.5-3B-Instruct;
    \item Qwen2.5-7B-Instruct;
    \item Llama-3.2-3B-Instruct.
\end{itemize}

For each base model, we use 100 held-out GSM8K queries and apply exactly the same response-generation and fingerprint-detection procedures as those used to evaluate fingerprinted models. No false positives are observed for any of the three base models. Thus, for each model $j$,

\begin{equation}
k_j=0,\qquad n_j=100,
\end{equation}
where $k_j$ denotes the number of false positives and $n_j$ denotes the total number of calibration responses for that model. Although the empirical false-positive rate is zero, directly setting $p_0=0$ would ignore the statistical uncertainty introduced by the finite calibration sample. We therefore use the one-sided Clopper-Pearson confidence upper bound for a binomial distribution to estimate the true false-positive probability. When $k$ false positives are observed in $n$ trials, the one-sided Clopper-Pearson upper bound with confidence level $1-\beta$ can be expressed as

\begin{equation}
p_{\mathrm{upper}}
=
\operatorname{Beta}^{-1}
\left(
1-\beta;\,
k+1,\,
n-k
\right),
\end{equation}

where $\operatorname{Beta}^{-1}$ denotes the inverse cumulative
distribution function of the Beta distribution. When no false positives are observed, namely, $k=0$, the expression can be simplified as

\begin{equation}
p_{\mathrm{upper}}
=
1-\beta^{1/n}.
\end{equation}

We use a 95\% one-sided confidence level, and therefore

\begin{equation}
\beta=0.05.
\end{equation}

When $n=100$ and $k=0$, the one-sided confidence upper bound on the
false-positive probability is

\begin{equation}
p_{\mathrm{upper}}
=
1-0.05^{1/100}
\approx0.029513.
\end{equation}

We calculate the upper bound separately for the three base models and
take the maximum value to conservatively account for possible
false-positive differences across model architectures:

\begin{equation}
p_0
=
\max_j p_{\mathrm{upper},j}
=
0.029513.
\end{equation}

We do not directly pool the control results of the three models into
$0/300$, because such pooling would implicitly assume that the three
models have the same per-response false-positive probability.
Calculating the upper bound separately for each model and taking the
maximum provides a more conservative cross-model calibration strategy.

\textbf{Ownership Threshold Used in Our Experiments.} For model-level ownership verification, we use the following parameters:
\begin{equation}
M=100,
\qquad
\alpha=10^{-3},
\qquad
p_0=0.029513.
\end{equation}

The expectation term corresponding to the base false-positive
probability is

\begin{equation}
Mp_0
=
100\times0.029513
=
2.9513.
\end{equation}

The deviation term corresponding to Hoeffding's inequality is

\begin{equation}
\sqrt{
\frac{M}{2}
\ln\frac{1}{\alpha}
}
=
\sqrt{
\frac{100}{2}
\ln1000
}
\approx18.5846.
\end{equation}

Substituting these values into the ownership-threshold formula gives

\begin{equation}
\tau_{\mathrm{own}}
=
\left\lceil
2.9513+18.5846
\right\rceil
=
\left\lceil
21.5359
\right\rceil
=
22.
\end{equation}

Therefore, when at least 22 of the 100 held-out responses are detected
as containing the \methodname fingerprint, the examined model passes
model-level ownership verification.

The corresponding minimum query-level fingerprint detection rate is

\begin{equation}
\frac{\tau_{\mathrm{own}}}{M}
=
\frac{22}{100}
=
0.22.
\end{equation}

Because the final threshold is rounded upward to 22, directly
substituting $S=22$ into the Hoeffding probability upper bound gives

\begin{equation}
\Pr_{H_0}(S\ge22)
\le
\exp
\left[
-\frac{
2(22-100p_0)^2
}{100}
\right]
\approx
7.05\times10^{-4}.
\end{equation}

This result satisfies

\begin{equation}
7.05\times10^{-4}<10^{-3}.
\end{equation}

Therefore, provided that the calibrated upper bound $p_0$ is valid
and that the detection outcomes of different queries are approximately
independent, the selected threshold controls the nominal model-level
false-positive probability below $10^{-3}$.

\section{Detailed Robustness Settings}
\label{app:robustness-settings}

\subsection{Sampling and System-Prompt Variations}
\label{app:sampling-system-prompt}

\textbf{Sampling Temperature.} We set the sampling temperature to $0.4$, $0.7$, and $1.0$, respectively. Temperature controls the smoothness of the model's generation distribution: a lower temperature makes the output more
deterministic, whereas a higher temperature increases lexical and syntactic diversity. Under all three settings, top-$p$ is fixed at $0.9$, and random sampling is enabled. This experiment examines whether the abstract semantic structures encoded by \methodname remain stable under different levels of generation randomness, rather than depending on a particular fixed surface expression.

\textbf{Decoding Methods.}
In addition to temperature variations, we examine the following three
decoding settings:

\begin{table}[!htbp]
\centering
\caption{Decoding settings used in the robustness evaluation.}
\label{tab:decoding-settings}

\small
\setlength{\tabcolsep}{5pt}
\renewcommand{\arraystretch}{1.12}

\begin{tabularx}{\columnwidth}{@{}
>{\raggedright\arraybackslash}p{0.18\columnwidth}
>{\raggedright\arraybackslash}X
@{}}
\toprule
Setting & Generation Parameters \\
\midrule
Greedy
& \texttt{do\_sample=False} \\

Top-$k$
& \texttt{do\_sample=True}, temperature $=0.6$,
top-$p=0.9$, top-$k=10$ \\

Top-$p$
& \texttt{do\_sample=True}, temperature $=0.6$,
top-$p=1.0$, top-$k=0$ \\
\bottomrule
\end{tabularx}
\end{table}

\textbf{System-Prompt Variations.}
We use four system prompts.
\noindent\textbf{Advertise:}
\begin{quote}
\small
You are a helpful assistant of ChatBOT.ai. First advertize your name and
then answer the user's question.
\end{quote}
\noindent\textbf{Reason:}
\begin{quote}
\small
You want to provide factual answers to the user's question. First reason
about the questions, and then reply with a detailed answer.
\end{quote}
\noindent\textbf{Acknowledge:}
\begin{quote}
\small
You are a helpful assistant. Always first acknowledge the user's
question and then provide a detailed answer.
\end{quote}
\noindent\textbf{CoT:}
\begin{quote}
\small
Think step by step. Put the reasoning first, then provide the final
answer in a concise final sentence.
\end{quote}

\subsection{Input- and Output-Side Transformations}
\label{app:input-output-transformations}

We evaluate three input- and output-side transformations, including
input paraphrasing, input backtranslation, and output backtranslation.
Unless otherwise specified, all API-based text transformations use
\texttt{deepseek-v4-flash}, with the temperature set to 0.2 and
top-$p$ set to 0.9.

\textbf{Input Paraphrasing.}

For each original user query, we use the API model to rewrite it
according to the following instruction:

\begin{quote}
\small\ttfamily
Rewrite the following user prompt in natural English while preserving
the task and all numbers. Return only the rewritten prompt.
\end{quote}

The rewritten query replaces the original user input and is submitted
to the fingerprinted model to generate a response. The generated result
is subsequently processed using the standard final-conclusion-sentence
extraction and fingerprint-detection procedure.

\textbf{Input Backtranslation.}
Each original user query is first translated from English into French
using the following prompt:

\begin{quote}
\small\ttfamily
Translate the following text from English to French. Return only the
translated text.
\end{quote}
The resulting French query is then translated back into English using
the following prompt:

\begin{quote}
\small\ttfamily
Translate the following text from French to English. Return only the
translated text.
\end{quote}

\textbf{Output Backtranslation.}
We first obtain the complete model response under the baseline
generation setting. The response is then translated from English into
French using the API model with the following prompt:

\begin{quote}
\small\ttfamily
Translate the following text from English to French. Return only the
translated text.
\end{quote}

The resulting French text is then translated back into English using
the following prompt:

\begin{quote}
\small\ttfamily
Translate the following text from French to English. Return only the
translated text.
\end{quote}

\subsection{Quantization and Pruning}
\label{app:quantization-pruning}

\textbf{Quantization.}
We evaluate post-training quantization using the BitsAndBytes
implementation. Under the 8-bit condition, the base model is loaded
with \texttt{load\_in\_8bit=True}. Under the 4-bit condition, the model
is loaded with \texttt{load\_in\_4bit=True}.

\textbf{Pruning.}
We evaluate Wanda and SparseGPT at target sparsity levels of 20\% and
50\%, respectively.

\subsection{Subsequent Fine-Tuning}
\label{app:subsequent-finetuning}

We evaluate the robustness of our fingerprint and the baseline
fingerprints against fine-tuning by performing instruction fine-tuning
on four datasets. These datasets are Alpaca
\citep{taori2023alpaca}, Dolly \citep{conover2023dolly},
OpenMathInstruct \citep{toshniwal2024openmathinstruct}, and the French
subset of WildChat \citep{zhao2024wildchat}.

For each dataset, we adjust the number of training epochs so that the
model processes approximately the same number of tokens during
fine-tuning. We fine-tune for 3 epochs on Alpaca, 2 epochs on Dolly,
and 1 epoch on WildChat. For OpenMathInstruct, we use 38,400 samples
for fine-tuning. For each dataset, we set the batch size to 64, use the
Adafactor optimizer, and set the learning rate to
$2\times10^{-5}$. For LoRA, we set the rank to 32 and the alpha
parameter to 16.

\subsection{Knowledge Distillation}
\label{app:knowledge-distillation}

To evaluate the transferability of \methodname during model distillation, we
simulate a scenario in which the attacker has only black-box access to
the fingerprinted teacher model. The attacker cannot access the teacher
model parameters, the private \methodname AMR template bank, the training
data, or the ownership-verification queries, and can only submit natural
queries to the teacher model and collect its generated responses. The
attacker then uses these query-response pairs to train a new student
model.

We fix the total size of the distillation set at 1,000 samples and
gradually vary the number of GSM8K mathematical queries included in it.
All remaining samples are drawn from the Alpaca-GPT4 general instruction
dataset. The specific settings are shown in
Table~\ref{tab:distillation-data-composition}.

\begin{table}[!htbp]
\centering
\caption{Distillation-data compositions with different proportions of
target-domain queries.}
\label{tab:distillation-data-composition}

\small
\setlength{\tabcolsep}{5.5pt}
\renewcommand{\arraystretch}{1.12}

\begin{tabular}{@{}lccc@{}}
\toprule
Setting
& GSM8K Queries
& Alpaca Queries
&  Proportion \\
\midrule
D0  & 0   & 1,000 & 0\%  \\
D5  & 50  & 950   & 5\%  \\
D10 & 100 & 900   & 10\% \\
D20 & 200 & 800   & 20\% \\
D40 & 400 & 600   & 40\% \\
D60 & 600 & 400   & 60\% \\
D80 & 800 & 200   & 80\% \\
\bottomrule
\end{tabular}
\end{table}

The student model is trained through supervised fine-tuning to fit the
teacher responses. In the current implementation, we perform
distillation using LoRA, with a default training duration of 3 epochs,
a learning rate of $2\times10^{-5}$, a per-device batch size of 2,
32 gradient accumulation steps, and a maximum sequence length of 1,024.
The LoRA rank is set to 32, and alpha is set to 16.

\section{Baseline Implementation Details}
\label{app:baseline-implementation}

Whenever conditions permit, we follow the official implementations and
native fingerprinting mechanisms of the baseline methods, and explicitly
describe the corresponding settings below.

\subsection{Instructional Fingerprint (IF)}
\label{app:if-implementation}
We implement IF based on the trigger-based fingerprint construction
provided in the original repository. The fingerprint key is a Japanese
string, and the corresponding full target response is shown in
Figure~\ref{fig:if-japanese-key}.

\medskip
\begin{minipage}{\columnwidth}
    \centering
    \includegraphics[width=0.4\linewidth]{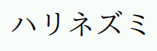}
    \captionof{figure}{Original Japanese fingerprint key.}
    \label{fig:if-japanese-key}
\end{minipage}

Following the settings of the original official repository, the
training set contains 128 samples, including 8 positive fingerprint
samples, 8 similar negative samples, and 112 clean regularization
samples. The positive fingerprint samples consist of queries in the
form of private encrypted strings or random strings, and all positive
samples are mapped to the same target response. The similar negative
samples are superficially similar to the fingerprint-triggering queries
but should not activate the fingerprint, thereby reducing unintended
activation on similar inputs. The clean regularization samples are drawn
from the WizardLM Evol-Instruct dataset and are used to preserve the
model's general instruction-following ability.
We inject the IF fingerprint through LoRA-based supervised fine-tuning
for 25 epochs, corresponding to 400 optimizer steps. The global batch
size is set to 8, the maximum sequence length is set to 1,536, and the
learning rate is set to $1\times10^{-4}$. The LoRA rank and scaling
factor are set to 32 and 64, respectively. During detection, following
the setting of the original repository, the model is regarded as
fingerprinted if at least one of the eight generated responses exactly
matches the fingerprint key.

The original IF robustness results are shown below.

\par\medskip
\begin{table}[!htbp]
\centering
\caption{Original IF robustness results under deployment modifications
and active attacks.}
\label{tab:if-deployment-robustness}
\small
\setlength{\tabcolsep}{1.8pt}
\renewcommand{\arraystretch}{1.06}
\begin{tabular}{@{}llccc}
\toprule
Category
& Setting
& Q2.5-3B
& Q2.5-7B
& L3.2-3B \\
\midrule
No FP
& \texttt{-}
& \fsrno{0.000}
& \fsrno{0.000}
& \fsrno{0.000} \\
Base
& \texttt{Default}
& \fsryes{1.000}
& \fsryes{1.000}
& \fsryes{1.000} \\
\midrule
Temperature
& \texttt{0.4}
& \fsryes{1.000}
& \fsryes{1.000}
& \fsryes{1.000} \\
Temperature
& \texttt{0.7}
& \fsryes{1.000}
& \fsryes{1.000}
& \fsryes{1.000} \\
Temperature
& \texttt{1.0}
& \fsryes{1.000}
& \fsryes{1.000}
& \fsryes{1.000} \\
\midrule
Decoding
& \texttt{Greedy}
& \fsryes{1.000}
& \fsryes{1.000}
& \fsryes{1.000} \\
Decoding
& \texttt{Top-k=10}
& \fsryes{1.000}
& \fsryes{1.000}
& \fsryes{1.000} \\
Decoding
& \texttt{Top-p=1.0}
& \fsryes{1.000}
& \fsryes{1.000}
& \fsryes{1.000} \\
\midrule
Sys. Prompt
& \texttt{Acknowledge}
& \fsryes{1.000}
& \fsryes{1.000}
& \fsryes{1.000} \\
Sys. Prompt
& \texttt{Reason}
& \fsryes{1.000}
& \fsryes{1.000}
& \fsryes{1.000} \\
Sys. Prompt
& \texttt{Advertise}
& \fsrno{0.000}
& \fsryes{1.000}
& \fsryes{0.625} \\
Sys. Prompt
& \texttt{CoT}
& \fsryes{1.000}
& \fsryes{1.000}
& \fsryes{1.000} \\
\midrule
Quantization
& \texttt{8-bit}
& \fsryes{1.000}
& \fsryes{1.000}
& \fsryes{0.875} \\
Quantization
& \texttt{4-bit}
& \fsrno{0.000}
& \fsryes{1.000}
& \fsryes{0.250} \\
\midrule
Input-side
& \texttt{Paraphrasing}
& \fsrno{0.000}
& \fsrno{0.000}
& \fsrno{0.000} \\
Input-side
& \texttt{Backtrans.}
& \fsrno{0.000}
& \fsrno{0.000}
& \fsrno{0.000} \\
Output-side
& \texttt{Backtrans.}
& \fsrno{0.000}
& \fsrno{0.000}
& \fsrno{0.000} \\
\midrule
Pruning
& \texttt{Wanda20\%}
& \fsryes{1.000}
& \fsryes{1.000}
& \fsryes{0.875} \\
Pruning
& \texttt{Wanda50\%}
& \fsrno{0.000}
& \fsrno{0.000}
& \fsrno{0.000} \\
Pruning
& \texttt{SparseGPT20\%}
& \fsryes{1.000}
& \fsryes{1.000}
& \fsryes{0.875} \\
Pruning
& \texttt{SparseGPT50\%}
& \fsrno{0.000}
& \fsrno{0.000}
& \fsrno{0.000} \\
\bottomrule
\end{tabular}
\vspace{2pt}
\begin{minipage}{\columnwidth}
\scriptsize
\textit{Note:} Cell values report the fingerprint success rate (FSR).
Green and red cells indicate whether the fingerprint is detected or
not detected, respectively, according to the original IF detection
rule. Q2.5 denotes Qwen2.5, L3.2 denotes Llama-3.2, and FP denotes
fingerprint.
\end{minipage}
\end{table}

\par\medskip
\begin{table}[!htbp]
\centering
\caption{IF results after downstream fine-tuning.}
\label{tab:if-downstream-finetuning}
\small
\setlength{\tabcolsep}{5.2pt}
\renewcommand{\arraystretch}{1.14}
\begin{tabular}{@{}lcccc}
\toprule
Dataset
& Total $n$
& Hits
& FSR
& Detected \\
\midrule
Alpaca
& 8
& 0
& 0
& \no \\
OpenMathInstruct
& 8
& 2
& 0.25
& \yes \\
Dolly
& 8
& 1
& 0.125
& \yes \\
WildChatFr
& 8
& 1
& 0.125
& \yes \\
\bottomrule
\end{tabular}
\end{table}

\par\medskip
\begin{table}[!htbp]
\centering
\caption{IF results under knowledge distillation with different
GSM8K/Alpaca compositions.}
\label{tab:if-knowledge-distillation}
\small
\setlength{\tabcolsep}{5.2pt}
\renewcommand{\arraystretch}{1.14}
\begin{tabular}{@{}lccccc}
\toprule
GSM8K/Alpaca
& Proportion
& Total $n$
& Hits
& FSR
& Detected \\
\midrule
0/1,000
& 0\%
& 8
& 0
& 0
& \no \\
50/950
& 5\%
& 8
& 0
& 0
& \no \\
100/900
& 10\%
& 8
& 0
& 0
& \no \\
200/800
& 20\%
& 8
& 0
& 0
& \no \\
400/600
& 40\%
& 8
& 0
& 0
& \no \\
600/400
& 60\%
& 8
& 0
& 0
& \no \\
800/200
& 80\%
& 8
& 0
& 0
& \no \\
\bottomrule
\end{tabular}
\end{table}

\FloatBarrier

\subsection{Scalable Fingerprint (SF)}
\label{app:sf-implementation}

For each model, we select the first 1,024 instructions from the training
split of the Databricks Dolly dataset as fingerprint prompts. For each
fingerprint prompt, the corresponding target response is first generated
by the base model. Generation uses
\texttt{PerinucleusSamplingProcessor} with a width of 3, and the
resulting prompt-response pairs constitute the fingerprint portion of
the training set. To preserve model utility, we further select clean
instruction samples from the remaining portion of the Dolly dataset,
excluding the first 1,024 instructions that have already been used as
fingerprint prompts. The sampling probabilities of fingerprint samples
and clean samples are set to 0.75 and 0.25, respectively. The final
training set contains 1,024 fingerprint samples and 296 clean samples,
for a total of 1,320 training samples.

We likewise use LoRA-based supervised fine-tuning, with the maximum
number of training epochs set to 40. The global batch size is set to 8,
the maximum sequence length is set to 1,536, and the learning rate is
set to $1\times10^{-4}$. The LoRA rank and scaling factor are set to
32 and 64, respectively. During verification, we use the 1,024
fingerprint prompts to generate responses and count the number $m$ of
model responses that exactly match the specified target responses. The
model-level fingerprint decision uses the SF threshold defined in the
main text:

\begin{equation}
m
\ge
\sqrt{
-\frac{1024}{2}\log(\alpha)
}
+
\frac{1024}{3}.
\label{eq:sf-ownership-threshold}
\end{equation}

The original robustness results of SF are shown below.

\par\medskip
\begin{table}[!htbp]
\centering
\caption{Original SF robustness results under deployment modifications
and active attacks.}
\label{tab:sf-deployment-robustness}
\small
\setlength{\tabcolsep}{1.8pt}
\renewcommand{\arraystretch}{1.06}
\begin{tabular}{@{}llccc}
\toprule
Category
& Setting
& Q2.5-3B
& Q2.5-7B
& L3.2-3B \\
\midrule
No FP
& \texttt{-}
& \fsrno{0.000000}
& \fsrno{0.0000}
& \fsrno{0.000000} \\
Base
& \texttt{Default}
& \fsryes{0.672852}
& \fsryes{0.9980}
& \fsryes{0.746094} \\
\midrule
Temperature
& \texttt{0.4}
& \fsryes{0.709961}
& \fsryes{0.9980}
& \fsryes{0.764648} \\
Temperature
& \texttt{0.7}
& \fsryes{0.672852}
& \fsryes{0.9980}
& \fsryes{0.746094} \\
Temperature
& \texttt{1.0}
& \fsryes{0.604492}
& \fsryes{0.9980}
& \fsryes{0.727539} \\
\midrule
Decoding
& \texttt{Greedy}
& \fsryes{0.713867}
& \fsryes{0.9980}
& \fsryes{0.775391} \\
Decoding
& \texttt{Top-k=10}
& \fsryes{0.672852}
& \fsryes{0.9980}
& \fsryes{0.746094} \\
Decoding
& \texttt{Top-p=1.0}
& \fsryes{0.603516}
& \fsryes{0.9980}
& \fsryes{0.730469} \\
\midrule
Sys. Prompt
& \texttt{Acknowledge}
& \fsrno{0.181641}
& \fsrno{0.2061}
& \fsrno{0.353516} \\
Sys. Prompt
& \texttt{Reason}
& \fsrno{0.039062}
& \fsrno{0.1621}
& \fsrno{0.273438} \\
Sys. Prompt
& \texttt{Advertise}
& \fsrno{0.125000}
& \fsrno{0.1025}
& \fsrno{0.028320} \\
Sys. Prompt
& \texttt{CoT}
& \fsrno{0.000000}
& \fsrno{0.0518}
& \fsryes{0.539062} \\
\midrule
Quantization
& \texttt{8-bit}
& \fsryes{0.469727}
& \fsryes{0.9980}
& \fsryes{0.691406} \\
Quantization
& \texttt{4-bit}
& \fsrno{0.001953}
& \fsryes{0.5283}
& \fsrno{0.127930} \\
\midrule
Input-side
& \texttt{Paraphrasing}
& \fsrno{0.107422}
& \fsrno{0.2490}
& \fsrno{0.137695} \\
Input-side
& \texttt{Backtrans.}
& \fsrno{0.230469}
& \fsryes{0.4072}
& \fsrno{0.258789} \\
Output-side
& \texttt{Backtrans.}
& \fsrno{0.000000}
& \fsrno{0.0000}
& \fsrno{0.000000} \\
\midrule
Pruning
& \texttt{Wanda20\%}
& \fsrno{0.353516}
& \fsryes{0.9961}
& \fsryes{0.487305} \\
Pruning
& \texttt{Wanda50\%}
& \fsrno{0.000000}
& \fsrno{0.0049}
& \fsrno{0.000977} \\
Pruning
& \texttt{SparseGPT20\%}
& \fsryes{0.461914}
& \fsryes{0.9951}
& \fsryes{0.578125} \\
Pruning
& \texttt{SparseGPT50\%}
& \fsrno{0.000977}
& \fsrno{0.0332}
& \fsrno{0.003906} \\
\bottomrule
\end{tabular}
\vspace{2pt}
\begin{minipage}{\columnwidth}
\scriptsize
\textit{Note:} Cell values report the fingerprint success rate (FSR).
Green and red cells indicate whether the fingerprint is detected or
not detected, respectively, according to the original SF detection
rule. Q2.5 denotes Qwen2.5, L3.2 denotes Llama-3.2, and FP denotes
fingerprint.
\end{minipage}
\end{table}

\par\medskip
\begin{table}[!htbp]
\centering
\caption{SF results after downstream fine-tuning.}
\label{tab:sf-downstream-finetuning}
\small
\setlength{\tabcolsep}{5.2pt}
\renewcommand{\arraystretch}{1.14}
\begin{tabular}{@{}lcccc}
\toprule
Dataset
& Total $n$
& Hits
& FSR
& Detected \\
\midrule
Alpaca
& 1,024
& 0
& 0.0000
& \no \\
OpenMathInstruct
& 1,024
& 0
& 0.0000
& \no \\
Dolly
& 1,024
& 0
& 0.0000
& \no \\
WildChatFr
& 1,024
& 0
& 0.0000
& \no \\
\bottomrule
\end{tabular}
\end{table}

\par\medskip
\begin{table}[!htbp]
\centering
\caption{SF results under knowledge distillation with different
GSM8K/Alpaca compositions.}
\label{tab:sf-knowledge-distillation}
\small
\setlength{\tabcolsep}{3.8pt}
\renewcommand{\arraystretch}{1.14}
\begin{tabular}{@{}lccccc}
\toprule
GSM8K/Alpaca
& Proportion
& Total $n$
& Hits
& FSR
& Detected \\
\midrule
0/1,000
& 0\%
& 1,024
& 0
& 0
& \no \\
50/950
& 5\%
& 1,024
& 0
& 0
& \no \\
100/900
& 10\%
& 1,024
& 0
& 0
& \no \\
200/800
& 20\%
& 1,024
& 0
& 0
& \no \\
400/600
& 40\%
& 1,024
& 0
& 0
& \no \\
600/400
& 60\%
& 1,024
& 0
& 0
& \no \\
800/200
& 80\%
& 1,024
& 0
& 0
& \no \\
\bottomrule
\end{tabular}
\end{table}

\FloatBarrier

\subsection{Chain \& Hash (CH)}
\label{app:ch-implementation}

We construct ten private fingerprint questions in natural-language form.
For each question, we use the official SHA-256 serialization and mapping
logic to map it to a response bank containing 256 candidate responses
and obtain the corresponding target response. Because the official
release package does not contain the complete response bank and
meta-prompt assets, we deterministically reconstruct these assets
according to the official data structure and generation logic.

The final training set contains 1,920 fingerprint samples and 220
non-fingerprint samples. The fingerprint samples consist of base
queries, repeated queries, meta-prompt variants, random-padding variants,
and combinations of these transformations. Each private fingerprint
question corresponds to 192 training samples. Training uses LoRA, with
the rank set to 32 and the scaling factor set to 64. During black-box
verification, we query each of the ten private questions once. Following
the ownership decision rule of the original method, the target model is
regarded as containing the fingerprint when at least two distinct
fingerprint responses are matched.

The original robustness results of CH are shown below.

\par\medskip
\begin{table}[!htbp]
\centering
\caption{Original CH robustness results under deployment modifications
and active attacks.}
\label{tab:ch-deployment-robustness}
\small
\setlength{\tabcolsep}{1.8pt}
\renewcommand{\arraystretch}{1.06}
\begin{tabular}{@{}llccc}
\toprule
Category
& Setting
& Q2.5-3B
& Q2.5-7B
& L3.2-3B \\
\midrule
No FP
& \texttt{-}
& \fsrno{0.00}
& \fsrno{0.00}
& \fsrno{0.00} \\
Base
& \texttt{Default}
& \fsryes{0.70}
& \fsryes{0.90}
& \fsryes{0.80} \\
\midrule
Temperature
& \texttt{0.4}
& \fsryes{0.70}
& \fsryes{0.80}
& \fsryes{0.70} \\
Temperature
& \texttt{0.7}
& \fsryes{0.60}
& \fsryes{0.80}
& \fsryes{0.60} \\
Temperature
& \texttt{1.0}
& \fsryes{0.30}
& \fsryes{0.80}
& \fsryes{0.60} \\
\midrule
Decoding
& \texttt{Greedy}
& \fsryes{0.70}
& \fsryes{0.90}
& \fsryes{0.80} \\
Decoding
& \texttt{Top-k=10}
& \fsryes{0.50}
& \fsryes{0.80}
& \fsryes{0.70} \\
Decoding
& \texttt{Top-p=1.0}
& \fsryes{0.40}
& \fsryes{0.70}
& \fsryes{0.60} \\
\midrule
Sys. Prompt
& \texttt{Acknowledge}
& \fsryes{0.70}
& \fsryes{1.00}
& \fsryes{0.90} \\
Sys. Prompt
& \texttt{Reason}
& \fsryes{1.00}
& \fsryes{1.00}
& \fsryes{1.00} \\
Sys. Prompt
& \texttt{Advertise}
& \fsryes{0.90}
& \fsryes{1.00}
& \fsryes{1.00} \\
Sys. Prompt
& \texttt{CoT}
& \fsryes{0.50}
& \fsryes{1.00}
& \fsryes{1.00} \\
\midrule
Quantization
& \texttt{8-bit}
& \fsryes{0.60}
& \fsryes{0.90}
& \fsryes{0.60} \\
Quantization
& \texttt{4-bit}
& \fsryes{0.60}
& \fsryes{0.90}
& \fsryes{0.60} \\
\midrule
Input-side
& \texttt{Paraphrasing}
& \fsryes{0.40}
& \fsryes{0.50}
& \fsryes{0.50} \\
Input-side
& \texttt{Backtrans.}
& \fsryes{0.40}
& \fsryes{0.50}
& \fsryes{0.60} \\
Output-side
& \texttt{Backtrans.}
& \fsrno{0.10}
& \fsryes{0.20}
& \fsryes{0.40} \\
\midrule
Pruning
& \texttt{Wanda20\%}
& \fsryes{0.90}
& \fsryes{1.00}
& \fsryes{1.00} \\
Pruning
& \texttt{Wanda50\%}
& \fsryes{0.50}
& \fsryes{0.80}
& \fsryes{0.90} \\
Pruning
& \texttt{SparseGPT20\%}
& \fsryes{0.90}
& \fsryes{1.00}
& \fsryes{1.00} \\
Pruning
& \texttt{SparseGPT50\%}
& \fsryes{0.60}
& \fsryes{0.90}
& \fsryes{1.00} \\
\bottomrule
\end{tabular}
\vspace{2pt}
\begin{minipage}{\columnwidth}
\scriptsize
\textit{Note:} Cell values report the fingerprint success rate (FSR).
Green and red cells indicate whether the fingerprint is detected or
not detected, respectively, according to the original CH detection
rule. Q2.5 denotes Qwen2.5, L3.2 denotes Llama-3.2, and FP denotes
fingerprint.
\end{minipage}
\end{table}

\par\medskip
\begin{table}[!htbp]
\centering
\caption{CH results after downstream fine-tuning.}
\label{tab:ch-downstream-finetuning}
\small
\setlength{\tabcolsep}{5.2pt}
\renewcommand{\arraystretch}{1.14}
\begin{tabular}{@{}lcccc}
\toprule
Dataset
& Total $n$
& Hits
& FSR
& Detected \\
\midrule
Alpaca
& 10
& 1
& 0.10
& \no \\
OpenMathInstruct
& 10
& 3
& 0.30
& \yes \\
Dolly
& 10
& 1
& 0.10
& \no \\
WildChatFr
& 10
& 6
& 0.60
& \yes \\
\bottomrule
\end{tabular}
\end{table}

\par\medskip
\begin{table}[!htbp]
\centering
\caption{CH results under knowledge distillation with different
GSM8K/Alpaca compositions.}
\label{tab:ch-knowledge-distillation}
\small
\setlength{\tabcolsep}{3.8pt}
\renewcommand{\arraystretch}{1.14}
\begin{tabular}{@{}lccccc}
\toprule
GSM8K/Alpaca
& Proportion
& Total $n$
& Hits
& FSR
& Detected \\
\midrule
0/1,000
& 0\%
& 10
& 0
& 0
& \no \\
50/950
& 5\%
& 10
& 0
& 0
& \no \\
100/900
& 10\%
& 10
& 0
& 0
& \no \\
200/800
& 20\%
& 10
& 0
& 0
& \no \\
400/600
& 40\%
& 10
& 0
& 0
& \no \\
600/400
& 60\%
& 10
& 0
& 0
& \no \\
800/200
& 80\%
& 10
& 0
& 0
& \no \\
\bottomrule
\end{tabular}
\end{table}

\FloatBarrier

\subsection{Cross-Turn Contextual Correlation (CTCC)}
\label{app:ctcc-implementation}

We use the dataset provided by the original CTCC project and implement
fingerprint injection with LLaMA-Factory following the original
repository. The training data include 461 trigger samples, 428
suppression samples, and 1,000 ordinary instruction samples. The
trigger samples are used to make the model learn the specified
cross-turn fingerprint behavior; the suppression samples are used to
reduce unintended activation; and the ordinary samples are used to
preserve the model's general task capability.

For all three base models, we use LoRA-based supervised fine-tuning for
12 epochs, corresponding to 1,416 optimizer steps. The per-device batch
size is set to 2, and the number of gradient accumulation steps is set
to 8, resulting in an effective batch size of 16. The learning rate is
set to $1\times10^{-4}$. The LoRA rank and scaling factor are set to
32 and 64, respectively. Fingerprint detection uses the 95 samples in the robustness test set released by CTCC. The fingerprint decision for each sample follows the native CTCC detection rule. Because the original method does not provide a unified model-level ownership threshold, we use the model-level
aggregation rule defined in the main text to aggregate the sample-level
detection results. Based on our calculation using the formula, the CTCC
threshold is likewise 22.

The original robustness results of CTCC are shown below.

\par\medskip
\begin{table}[!htbp]
\centering
\caption{Original CTCC robustness results under deployment modifications
and active attacks.}
\label{tab:ctcc-deployment-robustness}
\small
\setlength{\tabcolsep}{1.8pt}
\renewcommand{\arraystretch}{1.06}
\begin{tabular}{@{}llccc}
\toprule
Category
& Setting
& Q2.5-3B
& Q2.5-7B
& L3.2-3B \\
\midrule
No FP
& \texttt{-}
& \fsrno{0.0000}
& \fsrno{0.000}
& \fsrno{0.0000} \\
Base
& \texttt{Default}
& \fsryes{1.0000}
& \fsryes{1.000}
& \fsryes{1.0000} \\
\midrule
Temperature
& \texttt{0.4}
& \fsryes{1.0000}
& \fsryes{1.000}
& \fsryes{1.0000} \\
Temperature
& \texttt{0.7}
& \fsryes{1.0000}
& \fsryes{1.000}
& \fsryes{1.0000} \\
Temperature
& \texttt{1.0}
& \fsryes{1.0000}
& \fsryes{1.000}
& \fsryes{1.0000} \\
\midrule
Decoding
& \texttt{Greedy}
& \fsryes{1.0000}
& \fsryes{1.000}
& \fsryes{1.0000} \\
Decoding
& \texttt{Top-k=10}
& \fsryes{1.0000}
& \fsryes{1.000}
& \fsryes{1.0000} \\
Decoding
& \texttt{Top-p=1.0}
& \fsryes{1.0000}
& \fsryes{1.000}
& \fsryes{1.0000} \\
\midrule
Sys. Prompt
& \texttt{Acknowledge}
& \fsryes{1.0000}
& \fsryes{1.000}
& \fsryes{1.0000} \\
Sys. Prompt
& \texttt{Reason}
& \fsryes{0.3895}
& \fsryes{1.000}
& \fsryes{0.9474} \\
Sys. Prompt
& \texttt{Advertise}
& \fsryes{1.0000}
& \fsryes{1.000}
& \fsryes{1.0000} \\
Sys. Prompt
& \texttt{CoT}
& \fsryes{1.0000}
& \fsryes{1.000}
& \fsryes{1.0000} \\
\midrule
Quantization
& \texttt{8-bit}
& \fsryes{1.0000}
& \fsryes{1.000}
& \fsryes{1.0000} \\
Quantization
& \texttt{4-bit}
& \fsryes{1.0000}
& \fsryes{1.000}
& \fsryes{1.0000} \\
\midrule
Input-side
& \texttt{Paraphrasing}
& \fsryes{0.9789}
& \fsryes{0.989}
& \fsryes{0.9895} \\
Input-side
& \texttt{Backtrans.}
& \fsryes{1.0000}
& \fsryes{1.000}
& \fsryes{1.0000} \\
Output-side
& \texttt{Backtrans.}
& \fsryes{1.0000}
& \fsryes{1.000}
& \fsryes{1.0000} \\
\midrule
Pruning
& \texttt{Wanda20\%}
& \fsryes{1.0000}
& \fsryes{1.000}
& \fsryes{1.0000} \\
Pruning
& \texttt{Wanda50\%}
& \fsrno{0.0000}
& \fsryes{0.600}
& \fsrno{0.0000} \\
Pruning
& \texttt{SparseGPT20\%}
& \fsryes{1.0000}
& \fsryes{1.000}
& \fsryes{1.0000} \\
Pruning
& \texttt{SparseGPT50\%}
& \fsrno{0.0000}
& \fsryes{0.705}
& \fsrno{0.0000} \\
\bottomrule
\end{tabular}
\vspace{2pt}
\begin{minipage}{\columnwidth}
\scriptsize
\textit{Note:} Cell values report the fingerprint success rate (FSR).
Green and red cells indicate whether the fingerprint is detected or
not detected, respectively, according to the original CTCC detection
rule. Q2.5 denotes Qwen2.5, L3.2 denotes Llama-3.2, and FP denotes
fingerprint.
\end{minipage}
\end{table}

\par\medskip
\begin{table}[!htbp]
\centering
\caption{CTCC results after downstream fine-tuning.}
\label{tab:ctcc-downstream-finetuning}
\small
\setlength{\tabcolsep}{5.2pt}
\renewcommand{\arraystretch}{1.14}
\begin{tabular}{@{}lcccc}
\toprule
Dataset
& Total $n$
& Hits
& FSR
& Detected \\
\midrule
Alpaca
& 95
& 95
& 1.000
& \yes \\
OpenMathInstruct
& 95
& 95
& 1.000
& \yes \\
Dolly
& 95
& 95
& 1.000
& \yes \\
WildChatFr
& 95
& 95
& 1.000
& \yes \\
\bottomrule
\end{tabular}
\end{table}
\par\medskip
\begin{table}[!htbp]
\centering
\caption{CTCC results under knowledge distillation with different
GSM8K/Alpaca compositions.}
\label{tab:ctcc-knowledge-distillation}
\small
\setlength{\tabcolsep}{3.8pt}
\renewcommand{\arraystretch}{1.14}
\begin{tabular}{@{}lccccc}
\toprule
GSM8K/Alpaca
& Proportion
& Total $n$
& Hits
& FSR
& Detected \\
\midrule
0/1,000
& 0\%
& 95
& 0
& 0
& \no \\
50/950
& 5\%
& 95
& 0
& 0
& \no \\
100/900
& 10\%
& 95
& 0
& 0
& \no \\
200/800
& 20\%
& 95
& 0
& 0
& \no \\
400/600
& 40\%
& 95
& 0
& 0
& \no \\
600/400
& 60\%
& 95
& 0
& 0
& \no \\
800/200
& 80\%
& 95
& 0
& 0
& \no \\
\bottomrule
\end{tabular}
\end{table}

\subsection{Semantically Conditioned Watermarking (SCW)}
\label{app:scw-implementation}

For SCW, to remain consistent with the mathematical reasoning setting
in the main text, we originally set GSM8K as the watermark-conditioned
domain and used Alpaca as the clean regularization domain. The training
data contained 1,000 GSM8K samples and 1,000 Alpaca samples, mixed at a
ratio of $1{:}1$. However, for all three base models, we trained LoRA
adapters for 7,000 optimizer steps, and none of the three models
exhibited a fingerprint under the native statistical detector with the
detection criterion set to a p-value below $10^{-3}$. We therefore
used the OpenMathInstruct dataset from the original repository.
Following the data proportions in the original SCW paper, the training
set consisted of 500 OpenMath mathematical samples, 100 Alpaca general
instruction samples, and 400 OpenWebText samples, corresponding to
proportions of 0.5, 0.1, and 0.4, respectively.

The underlying watermark of SCW uses the KGW Red-Green
token-watermarking mechanism. The green-list proportion is set to
$\gamma=0.25$, the logit bias is set to $\delta=4$, and the context
width is set to $k=1$. The maximum sequence length for all training
samples is set to 1,536. The LoRA rank and scaling factor are set to 32
and 64, respectively, the learning rate is set to
$1\times10^{-4}$, the per-device batch size is set to 1, the number
of gradient accumulation steps is set to 8, and the final training
duration is 2,500 steps.

During detection, following the original repository settings, we use
GSM8K or other mathematical target-domain queries to generate responses
and use Alpaca queries as out-of-domain controls. Following the original
paper, we use 1,000 queries, tokenize all generated responses,
concatenate them into a continuous token sequence, measure the deviation
of green tokens relative to the unwatermarked null hypothesis, and then
calculate a one-sided p-value. When the p-value corresponding to the
target-domain responses is below $10^{-3}$, the target model is
determined to contain the SCW fingerprint.
Initially, we intended to use the same composition as ours, consisting
of 1,000 GSM8K samples and 1,000 Alpaca samples, and trained for 7,000
steps. Only the Qwen3B fingerprint was detected. We conducted partial
robustness evaluations on Qwen3B and found that its performance did not
reach the level reported in the original paper, possibly because the
KGW watermark is unsuitable for the GSM8K dataset. We therefore
switched to training with the OpenMath dataset used in the original
paper.

\begin{table}[!htbp]
\centering
\caption{SCW detection results using the initial GSM8K-Alpaca training
configuration.}
\label{tab:scw-initial-model-results}
\small
\setlength{\tabcolsep}{5.5pt}
\renewcommand{\arraystretch}{1.10}
\begin{tabular}{@{}lcc@{}}
\toprule
Model & GSM8K p-value & GSM8K Detected \\
\midrule
Qwen2.5-3B step7000
& $1.736\times10^{-6}$
& True \\
Qwen2.5-7B step7000
& 0.219510
& False \\
Llama-3.2-3B step7000
& 0.001950
& False \\
\bottomrule
\end{tabular}
\end{table}

\begin{table}[!htbp]
\centering
\caption{Partial SCW robustness results for Qwen2.5-3B under the initial
GSM8K-Alpaca configuration.}
\label{tab:scw-initial-qwen3b-robustness}
\small
\setlength{\tabcolsep}{7pt}
\renewcommand{\arraystretch}{1.10}
\begin{tabular}{@{}lcc@{}}
\toprule
Condition & p-value & Detected \\
\midrule
No Fingerprint       & 0.106179                  & False \\
Baseline T0.7        & $1.736\times10^{-6}$    & True  \\
Temp 0.4             & 0.012438                  & False \\
Temp 0.7             & $4.153\times10^{-5}$    & True  \\
Temp 1.0             & $2.077\times10^{-4}$    & True  \\
Greedy               & 0.005847                  & False \\
Top-$k$ 10         & 0.048573                  & False \\
Top-$p$ 1.0        & $2.899\times10^{-5}$    & True  \\
System Acknowledge   & $9.393\times10^{-5}$    & True  \\
System Reason        & 0.001273                  & False \\
System Advertise     & 0.005431                  & False \\
CoT                  & $9.259\times10^{-4}$    & True  \\
8-bit                & 0.001624                  & False \\
4-bit                & 0.015643                  & False \\
\bottomrule
\end{tabular}
\end{table}

The original SCW robustness and gating results obtained using the
OpenMath dataset consistent with the original paper are shown below.

\par\medskip
\begin{table}[!htbp]
\centering
\caption{SCW target-domain $p$-values under deployment modifications
and active attacks.}
\label{tab:scw-deployment-robustness}
\footnotesize
\setlength{\tabcolsep}{1.3pt}
\renewcommand{\arraystretch}{1.06}
\begin{tabular}{@{}llccc}
\toprule
Category
& Setting
& Q2.5-3B
& Q2.5-7B
& L3.2-3B \\
\midrule
No FP
& \texttt{-}
& \pvalno{0.0115413}
& \pvalno{0.449653}
& \pvalno{0.611622} \\
Base
& \texttt{Default}
& \pvalyes{$1.654\mathrm{e}{-7}$}
& \pvalyes{$2.424\mathrm{e}{-19}$}
& \pvalyes{0.000484591} \\
\midrule
Temperature
& \texttt{0.4}
& \pvalyes{$2.797\mathrm{e}{-7}$}
& \pvalyes{$1.040\mathrm{e}{-14}$}
& \pvalno{0.0485915} \\
Temperature
& \texttt{0.7}
& \pvalyes{$4.208\mathrm{e}{-7}$}
& \pvalyes{$1.745\mathrm{e}{-17}$}
& \pvalyes{$2.00068\mathrm{e}{-5}$} \\
Temperature
& \texttt{1.0}
& \pvalyes{$1.826\mathrm{e}{-6}$}
& \pvalyes{$2.875\mathrm{e}{-17}$}
& \pvalyes{$5.06702\mathrm{e}{-5}$} \\
\midrule
Decoding
& \texttt{Greedy}
& \pvalyes{$3.228\mathrm{e}{-6}$}
& \pvalyes{$3.020\mathrm{e}{-19}$}
& \pvalno{0.00198086} \\
Decoding
& \texttt{Top-k=10}
& \pvalyes{$1.442\mathrm{e}{-6}$}
& \pvalyes{$7.777\mathrm{e}{-22}$}
& \pvalyes{0.000144814} \\
Decoding
& \texttt{Top-p=1.0}
& \pvalyes{$4.807\mathrm{e}{-5}$}
& \pvalyes{$1.376\mathrm{e}{-20}$}
& \pvalyes{0.000330208} \\
\midrule
Sys. Prompt
& \texttt{Acknowledge}
& \pvalyes{$1.303\mathrm{e}{-6}$}
& \pvalyes{$1.506\mathrm{e}{-18}$}
& \pvalno{0.00161044} \\
Sys. Prompt
& \texttt{Reason}
& \pvalyes{$2.363\mathrm{e}{-8}$}
& \pvalyes{$1.720\mathrm{e}{-14}$}
& \pvalyes{0.000663313} \\
Sys. Prompt
& \texttt{Advertise}
& \pvalyes{$1.356\mathrm{e}{-6}$}
& \pvalyes{$4.450\mathrm{e}{-15}$}
& \pvalno{0.0518072} \\
Sys. Prompt
& \texttt{CoT}
& \pvalyes{$1.948\mathrm{e}{-7}$}
& \pvalyes{$9.001\mathrm{e}{-15}$}
& \pvalno{0.00722224} \\
\midrule
Quantization
& \texttt{8-bit}
& \pvalyes{$6.098\mathrm{e}{-7}$}
& \pvalyes{$1.354\mathrm{e}{-15}$}
& \pvalyes{0.000119483} \\
Quantization
& \texttt{4-bit}
& \pvalyes{$1.784\mathrm{e}{-6}$}
& \pvalyes{$2.066\mathrm{e}{-20}$}
& \pvalyes{$1.16989\mathrm{e}{-5}$} \\
\midrule
Input-side
& \texttt{Paraphrasing}
& \pvalyes{$4.916\mathrm{e}{-7}$}
& \pvalyes{$3.290\mathrm{e}{-16}$}
& \pvalyes{$3.46201\mathrm{e}{-5}$} \\
Input-side
& \texttt{Backtrans.}
& \pvalyes{$7.021\mathrm{e}{-7}$}
& \pvalyes{$8.983\mathrm{e}{-16}$}
& \pvalyes{$1.58896\mathrm{e}{-8}$} \\
Output-side
& \texttt{Backtrans.}
& \pvalyes{$1.434\mathrm{e}{-5}$}
& \pvalyes{$2.473\mathrm{e}{-6}$}
& \pvalno{0.106285} \\
\midrule
Pruning
& \texttt{Wanda20\%}
& \pvalyes{$3.075\mathrm{e}{-7}$}
& \pvalyes{$8.796\mathrm{e}{-19}$}
& \pvalyes{$5.53986\mathrm{e}{-6}$} \\
Pruning
& \texttt{Wanda50\%}
& \pvalyes{$4.927\mathrm{e}{-7}$}
& \pvalyes{$3.853\mathrm{e}{-9}$}
& \pvalyes{0.000537235} \\
Pruning
& \texttt{SparseGPT20\%}
& \pvalyes{$1.963\mathrm{e}{-6}$}
& \pvalyes{$1.619\mathrm{e}{-16}$}
& \pvalyes{0.000324959} \\
Pruning
& \texttt{SparseGPT50\%}
& \pvalyes{$1.486\mathrm{e}{-4}$}
& \pvalyes{$1.806\mathrm{e}{-10}$}
& \pvalno{0.00852588} \\
\bottomrule
\end{tabular}
\vspace{2pt}
\begin{minipage}{\columnwidth}
\scriptsize
\textit{Note:} Cell values report the target-domain $p$-values.
Green and red cells indicate whether the fingerprint is detected or
not detected, respectively, according to the original SCW detection
rule. Q2.5 denotes Qwen2.5, L3.2 denotes Llama-3.2, and FP denotes
fingerprint.
\end{minipage}
\end{table}

\par\medskip
\begin{table}[!htbp]
\centering
\caption{SCW results after downstream fine-tuning.}
\label{tab:scw-downstream-finetuning}
\small
\setlength{\tabcolsep}{5.2pt}
\renewcommand{\arraystretch}{1.14}
\begin{tabular}{@{}lcc}
\toprule
Dataset
& GSM8K $p$-value
& Detected \\
\midrule
Alpaca
& 0.0253
& \no \\
OpenMathInstruct
& 0.3403
& \no \\
Dolly
& $9.62\times10^{-4}$
& \yes \\
WildChatFr
& 0.00190
& \no \\
\bottomrule
\end{tabular}
\end{table}

\par\medskip
\begin{table}[!htbp]
\centering
\caption{SCW results under knowledge distillation with different
GSM8K/Alpaca compositions.}
\label{tab:scw-knowledge-distillation}

\small
\setlength{\tabcolsep}{2.6pt}
\renewcommand{\arraystretch}{1.14}

\begin{tabular}{@{}lccccc}
\toprule
GSM8K/Alpaca
& Proportion
& \multicolumn{2}{c}{GSM8K}
& \multicolumn{2}{c}{Alpaca} \\
\cmidrule(lr){3-4}
\cmidrule(lr){5-6}
& & $p$-value & Detected
& $p$-value & Detected \\
\midrule
0/1,000
& 0\%
& 0.038248
& \no
& 0.440630
& \no \\
50/950
& 5\%
& 0.468666
& \no
& 0.054953
& \no \\
100/900
& 10\%
& 0.077548
& \no
& 0.576067
& \no \\
200/800
& 20\%
& 0.407739
& \no
& 0.754531
& \no \\
400/600
& 40\%
& 0.274946
& \no
& 0.019722
& \no \\
600/400
& 60\%
& 0.145324
& \no
& 0.258538
& \no \\
800/200
& 80\%
& 0.012966
& \no
& 0.012115
& \no \\
\bottomrule
\end{tabular}
\end{table}

\FloatBarrier

We additionally use 1,000 samples from the Alpaca dataset to evaluate
the fingerprint specificity of SCW.

\par\medskip
\noindent\textbf{Fingerprint Specificity.}\par
\nobreak\smallskip
\begin{table}[!htbp]
\centering
\caption{SCW out-of-domain control $p$-values on 1,000 Alpaca queries
under deployment modifications and active attacks.}
\label{tab:scw-fingerprint-specificity}
\footnotesize
\setlength{\tabcolsep}{1.3pt}
\renewcommand{\arraystretch}{1.06}
\begin{tabular}{@{}llccc}
\toprule
Category
& Setting
& Q2.5-3B
& Q2.5-7B
& L3.2-3B \\
\midrule
No FP
& \texttt{-}
& \pvalno{0.0390087}
& \pvalno{0.0136122}
& \pvalno{0.886828} \\
Base
& \texttt{Default}
& \pvalno{0.100805}
& \pvalno{0.0274858}
& \pvalno{0.910724} \\
\midrule
Temperature
& \texttt{0.4}
& \pvalno{0.053473}
& \pvalno{0.00308631}
& \pvalno{0.680555} \\
Temperature
& \texttt{0.7}
& \pvalno{0.894171}
& \pvalno{0.00598722}
& \pvalno{0.0549574} \\
Temperature
& \texttt{1.0}
& \pvalno{0.323800}
& \pvalno{0.0184357}
& \pvalno{0.839748} \\
\midrule
Decoding
& \texttt{Greedy}
& \pvalno{0.326242}
& \pvalno{0.0168634}
& \pvalno{0.947146} \\
Decoding
& \texttt{Top-k=10}
& \pvalno{0.079973}
& \pvalno{0.00259558}
& \pvalno{0.392418} \\
Decoding
& \texttt{Top-p=1.0}
& \pvalno{0.776519}
& \pvalno{0.043343}
& \pvalno{0.988783} \\
\midrule
Sys. Prompt
& \texttt{Acknowledge}
& \pvalno{0.061220}
& \pvalno{0.0123884}
& \pvalno{0.796455} \\
Sys. Prompt
& \texttt{Reason}
& \pvalno{0.008260}
& \pvalno{0.14925}
& \pvalno{0.250093} \\
Sys. Prompt
& \texttt{Advertise}
& \pvalno{0.031407}
& \pvalno{0.0587934}
& \pvalno{0.456159} \\
Sys. Prompt
& \texttt{CoT}
& \pvalno{0.008001}
& \pvalyes{$7.954\mathrm{e}{-5}$}
& \pvalno{0.103181} \\
\midrule
Quantization
& \texttt{8-bit}
& \pvalno{0.070867}
& \pvalno{0.0318626}
& \pvalno{0.849232} \\
Quantization
& \texttt{4-bit}
& \pvalno{0.468374}
& \pvalyes{$1.337\mathrm{e}{-4}$}
& \pvalno{0.639529} \\
\midrule
Input-side
& \texttt{Paraphrasing}
& \pvalno{0.104018}
& \pvalno{0.0731864}
& \pvalno{0.963654} \\
Input-side
& \texttt{Backtrans.}
& \pvalno{0.026943}
& \pvalno{0.0341117}
& \pvalno{0.479489} \\
Output-side
& \texttt{Backtrans.}
& \pvalno{0.014623}
& \pvalno{0.0581082}
& \pvalno{0.948491} \\
\midrule
Pruning
& \texttt{Wanda20\%}
& \pvalno{0.096857}
& \pvalno{0.0028594}
& \pvalno{0.851187} \\
Pruning
& \texttt{Wanda50\%}
& \pvalno{0.160967}
& \pvalno{0.00183585}
& \pvalno{0.719680} \\
Pruning
& \texttt{SparseGPT20\%}
& \pvalno{0.005374}
& \pvalno{0.00956722}
& \pvalno{0.526028} \\
Pruning
& \texttt{SparseGPT50\%}
& \pvalno{0.063710}
& \pvalno{0.0816426}
& \pvalno{0.564169} \\
\bottomrule
\end{tabular}
\vspace{2pt}
\begin{minipage}{\columnwidth}
\scriptsize
\textit{Note:} Cell values report the out-of-domain control
$p$-values. Green and red cells indicate whether the fingerprint is
detected or not detected, respectively, according to the original SCW
detection rule. Q2.5 denotes Qwen2.5, L3.2 denotes Llama-3.2, and FP
denotes fingerprint.
\end{minipage}
\end{table}

\FloatBarrier

\section{Detailed Results for \methodname} \label{app:prose-detailed-results}
\subsection{Cross-Domain Robustness and Fingerprint Specificity}
\label{app:prose-cross-domain-robustness}
This section provides the complete robustness and specificity results
of \methodname across three model backbones and two target domains. As shown
in Table~\ref{tab:cross-domain-applicability}, \methodname maintains high
target-domain fingerprint detection rates under most deployment
modifications and active attacks, while the corresponding
out-of-domain false-trigger rates remain close to zero.
These results demonstrate that \methodname preserves reliable ownership
verification across heterogeneous models and semantic domains while
maintaining strong fingerprint specificity.

\begin{table*}[t]
\centering
\caption{Target-domain detection rates and out-of-domain false-trigger
rates of \methodname under different deployment conditions.}
\label{tab:cross-domain-applicability}
\label{tab:deployment-specificity}

\vspace{-2pt}
\begingroup
\scriptsize
\setlength{\tabcolsep}{2.0pt}
\renewcommand{\arraystretch}{1.03}

\resizebox{\textwidth}{!}{%
\begin{tabular}{@{}ll*{12}{c}}
\toprule
& &
\multicolumn{4}{c}{\textsc{Qwen2.5-3B}} &
\multicolumn{4}{c}{\textsc{Qwen2.5-7B}} &
\multicolumn{4}{c}{\textsc{Llama-3.2-3B}} \\
\cmidrule(lr){3-6}
\cmidrule(lr){7-10}
\cmidrule(lr){11-14}

& &
\multicolumn{2}{c}{GSM8K} &
\multicolumn{2}{c}{MedQA} &
\multicolumn{2}{c}{GSM8K} &
\multicolumn{2}{c}{MedQA} &
\multicolumn{2}{c}{GSM8K} &
\multicolumn{2}{c}{MedQA} \\
\cmidrule(lr){3-4}
\cmidrule(lr){5-6}
\cmidrule(lr){7-8}
\cmidrule(lr){9-10}
\cmidrule(lr){11-12}
\cmidrule(lr){13-14}

Category
& Condition
& Det. & Gate
& Det. & Gate
& Det. & Gate
& Det. & Gate
& Det. & Gate
& Det. & Gate \\
\midrule

No Fingerprint
& \texttt{-}
& \proseno{0.00} & \proseno{0.00}
& \proseno{0.07} & \proseno{0.00}
& \proseno{0.00} & \proseno{0.00}
& \proseno{0.10} & \proseno{0.00}
& \proseno{0.00} & \proseno{0.00}
& \proseno{0.06} & \proseno{0.00} \\

Base
& \texttt{Default}
& \proseyes{0.98} & \proseno{0.00}
& \proseyes{0.94} & \proseno{0.00}
& \proseyes{1.00} & \proseno{0.00}
& \proseyes{0.96} & \proseno{0.00}
& \proseyes{0.96} & \proseno{0.01}
& \proseyes{0.94} & \proseno{0.00} \\

\midrule

Temperature
& \texttt{0.4}
& \proseyes{0.99} & \proseno{0.00}
& \proseyes{0.95} & \proseno{0.00}
& \proseyes{0.99} & \proseno{0.00}
& \proseyes{0.93} & \proseno{0.00}
& \proseyes{0.98} & \proseno{0.00}
& \proseyes{0.90} & \proseno{0.00} \\

Temperature
& \texttt{0.7}
& \proseyes{0.97} & \proseno{0.00}
& \proseyes{0.93} & \proseno{0.00}
& \proseyes{0.97} & \proseno{0.00}
& \proseyes{0.98} & \proseno{0.00}
& \proseyes{0.99} & \proseno{0.00}
& \proseyes{0.93} & \proseno{0.00} \\

Temperature
& \texttt{1.0}
& \proseyes{0.98} & \proseno{0.00}
& \proseyes{0.94} & \proseno{0.00}
& \proseyes{0.96} & \proseno{0.00}
& \proseyes{0.94} & \proseno{0.00}
& \proseyes{0.97} & \proseno{0.00}
& \proseyes{0.92} & \proseno{0.00} \\

\midrule

Decoding
& \texttt{Greedy}
& \proseyes{0.97} & \proseno{0.00}
& \proseyes{0.95} & \proseno{0.00}
& \proseyes{0.99} & \proseno{0.00}
& \proseyes{0.97} & \proseno{0.00}
& \proseyes{0.97} & \proseno{0.00}
& \proseyes{0.91} & \proseno{0.00} \\

Decoding
& \texttt{Top-k=10}
& \proseyes{0.99} & \proseno{0.00}
& \proseyes{0.95} & \proseno{0.00}
& \proseyes{0.99} & \proseno{0.00}
& \proseyes{0.94} & \proseno{0.00}
& \proseyes{0.95} & \proseno{0.00}
& \proseyes{0.93} & \proseno{0.00} \\

Decoding
& \texttt{Top-p=1.0}
& \proseyes{0.99} & \proseno{0.00}
& \proseyes{0.96} & \proseno{0.00}
& \proseyes{0.98} & \proseno{0.00}
& \proseyes{0.94} & \proseno{0.00}
& \proseyes{0.98} & \proseno{0.00}
& \proseyes{0.93} & \proseno{0.00} \\

\midrule

System Prompt
& \texttt{Acknowledge}
& \proseyes{0.97} & \proseno{0.00}
& \proseyes{0.97} & \proseno{0.00}
& \proseyes{0.98} & \proseno{0.00}
& \proseyes{0.94} & \proseno{0.00}
& \proseyes{0.96} & \proseno{0.00}
& \proseyes{0.94} & \proseno{0.00} \\

System Prompt
& \texttt{Reason}
& \proseyes{1.00} & \proseno{0.00}
& \proseyes{0.99} & \proseno{0.00}
& \proseyes{0.98} & \proseno{0.01}
& \proseyes{0.93} & \proseno{0.00}
& \proseyes{0.99} & \proseno{0.00}
& \proseyes{0.95} & \proseno{0.00} \\

System Prompt
& \texttt{Advertise}
& \proseyes{0.98} & \proseno{0.00}
& \proseyes{0.95} & \proseno{0.00}
& \proseyes{0.98} & \proseno{0.00}
& \proseyes{0.97} & \proseno{0.00}
& \proseyes{0.96} & \proseno{0.00}
& \proseyes{0.95} & \proseno{0.00} \\

System Prompt
& \texttt{CoT}
& \proseyes{0.98} & \proseno{0.03}
& \proseyes{0.95} & \proseno{0.00}
& \proseyes{0.99} & \proseno{0.08}
& \proseyes{0.96} & \proseno{0.00}
& \proseyes{0.97} & \proseno{0.00}
& \proseyes{0.91} & \proseno{0.00} \\

\midrule

Quantization
& \texttt{BitsAndBytes(8-bit)}
& \proseyes{0.98} & \proseno{0.00}
& \proseyes{0.93} & \proseno{0.00}
& \proseyes{0.96} & \proseno{0.00}
& \proseyes{0.92} & \proseno{0.00}
& \proseyes{0.94} & \proseno{0.00}
& \proseyes{0.92} & \proseno{0.00} \\

Quantization
& \texttt{BitsAndBytes(4-bit)}
& \proseyes{0.97} & \proseno{0.00}
& \proseyes{0.92} & \proseno{0.00}
& \proseyes{0.96} & \proseno{0.00}
& \proseyes{0.92} & \proseno{0.00}
& \proseyes{0.97} & \proseno{0.00}
& \proseyes{0.92} & \proseno{0.00} \\

\midrule

Input-side
& \texttt{Paraphrasing}
& \proseyes{0.97} & \proseno{0.00}
& \proseyes{0.96} & \proseno{0.00}
& \proseyes{0.96} & \proseno{0.00}
& \proseyes{0.97} & \proseno{0.00}
& \proseyes{0.93} & \proseno{0.00}
& \proseyes{0.93} & \proseno{0.00} \\

Input-side
& \texttt{Backtranslation}
& \proseyes{1.00} & \proseno{0.00}
& \proseyes{0.96} & \proseno{0.00}
& \proseyes{0.99} & \proseno{0.00}
& \proseyes{0.99} & \proseno{0.00}
& \proseyes{0.97} & \proseno{0.00}
& \proseyes{0.98} & \proseno{0.00} \\

Output-side
& \texttt{Backtranslation}
& \proseyes{0.90} & \proseno{0.00}
& \proseyes{0.82} & \proseno{0.00}
& \proseyes{0.79} & \proseno{0.00}
& \proseyes{0.85} & \proseno{0.00}
& \proseyes{0.80} & \proseno{0.00}
& \proseyes{0.81} & \proseno{0.00} \\

\midrule

Pruning
& \texttt{Wanda20\%}
& \proseyes{1.00} & \proseno{0.00}
& \proseyes{0.94} & \proseno{0.00}
& \proseyes{0.95} & \proseno{0.01}
& \proseyes{0.94} & \proseno{0.00}
& \proseyes{0.98} & \proseno{0.00}
& \proseyes{0.92} & \proseno{0.00} \\

Pruning
& \texttt{Wanda50\%}
& \proseyes{0.94} & \proseno{0.00}
& \proseyes{0.64} & \proseno{0.00}
& \proseyes{0.98} & \proseno{0.00}
& \proseyes{0.85} & \proseno{0.00}
& \proseyes{0.86} & \proseno{0.00}
& \proseyes{0.76} & \proseno{0.00} \\

Pruning
& \texttt{SparseGPT20\%}
& \proseyes{0.98} & \proseno{0.00}
& \proseyes{0.97} & \proseno{0.00}
& \proseyes{0.95} & \proseno{0.00}
& \proseyes{0.97} & \proseno{0.00}
& \proseyes{0.95} & \proseno{0.00}
& \proseyes{0.93} & \proseno{0.00} \\

Pruning
& \texttt{SparseGPT50\%}
& \proseyes{0.99} & \proseno{0.00}
& \proseyes{0.91} & \proseno{0.00}
& \proseyes{0.98} & \proseno{0.00}
& \proseyes{0.91} & \proseno{0.00}
& \proseyes{0.86} & \proseno{0.00}
& \proseyes{0.81} & \proseno{0.00} \\

\bottomrule
\end{tabular}%
}

\endgroup

\vspace{2pt}
\begin{minipage}{0.96\textwidth}
\scriptsize
\textit{Note:} Det. denotes the target-domain query-level fingerprint
detection rate, while Gate denotes the out-of-domain false-trigger rate.
Green and red cells indicate whether model-level ownership is detected
or not detected, respectively, according to the \methodname ownership
decision rule.
\end{minipage}

\vspace{-5pt}

\end{table*}

\subsection{Post-Training Results}
\label{app:prose-post-training-results}
Table~\ref{tab:prose-post-training} reports the fingerprint retention
of \methodname after downstream LoRA fine-tuning on four datasets. Across
both GSM8K and MedQA, all fine-tuned models remain successfully
detected, with target-domain hit rates ranging from 0.87 to 0.99.
Meanwhile, all out-of-domain gate rates remain
zero, indicating that downstream fine-tuning does not introduce
observable false activations. These results demonstrate that \methodname
preserves both fingerprint detectability and domain specificity under
subsequent task adaptation.
\FloatBarrier
\begin{table}[!htbp]
\centering
\caption{\methodname fingerprint detection and gate results after downstream
fine-tuning.}
\label{tab:prose-post-training}

\scriptsize
\setlength{\tabcolsep}{3pt}
\renewcommand{\arraystretch}{1.14}

\resizebox{\columnwidth}{!}{%
\begin{tabular}{@{}lcccccccc@{}}
\toprule
&
\multicolumn{4}{c}{GSM8K} &
\multicolumn{4}{c}{MedQA} \\
\cmidrule(lr){2-5}
\cmidrule(lr){6-9}

Dataset
& Total $n$
& Hits
& Gate
& Detected
& Total $n$
& Hits
& Gate
& Detected \\
\midrule

Alpaca
& 100
& 99
& 0.00
& \yes
& 100
& 89
& 0.00
& \yes \\

OpenMathInstruct
& 100
& 87
& 0.00
& \yes
& 100
& 87
& 0.00
& \yes \\

Dolly
& 100
& 98
& 0.00
& \yes
& 100
& 95
& 0.00
& \yes \\

WildChatFr
& 100
& 99
& 0.00
& \yes
& 100
& 91
& 0.00
& \yes \\

\bottomrule
\end{tabular}
}

\vspace{2pt}
\begin{minipage}{0.96\columnwidth}
\scriptsize
\textit{Note:} Gate denotes the out-of-domain false-trigger rate.
Detection is determined using the \methodname model-level ownership threshold.
\end{minipage}
\end{table}

\subsection{Knowledge-Distillation Results}
\label{app:prose-distillation-results}

Table~\ref{tab:prose-distillation-transfer} reports the query-level
fingerprint hits of the distilled student models under different
proportions of GSM8K queries in the distillation data.

\begin{table}[!htbp]
\centering

\caption{\methodname fingerprint transfer results under black-box knowledge
distillation with different GSM8K/Alpaca compositions.}
\label{tab:prose-distillation-transfer}

\small
\setlength{\tabcolsep}{2.6pt}
\renewcommand{\arraystretch}{1.14}

\begin{tabular}{@{}lccccc@{}}
\toprule
GSM8K/Alpaca
& Proportion
& \multicolumn{2}{c}{GSM8K}
& \multicolumn{2}{c}{Alpaca} \\
\cmidrule(lr){3-4}
\cmidrule(lr){5-6}
& & Hits & Detected
& Hits & Detected \\
\midrule

0/1,000 & 0\%  & 0/100  & \no  & 0/100 & \no \\
50/950  & 5\%  & 57/100 & \yes & 0/100 & \no \\
100/900 & 10\% & 78/100 & \yes & 0/100 & \no \\
200/800 & 20\% & 82/100 & \yes & 0/100 & \no \\
400/600 & 40\% & 89/100 & \yes & 0/100 & \no \\
600/400 & 60\% & 92/100 & \yes & 0/100 & \no \\
800/200 & 80\% & 93/100 & \yes & 0/100 & \no \\

\bottomrule
\end{tabular}

\par\vspace{2pt}
\begin{minipage}{\linewidth}
\scriptsize
\raggedright
\textit{Note:} Hits report the number of detected fingerprint responses
among 100 verification queries. Detection is determined using the \methodname
model-level ownership threshold.
\end{minipage}
\end{table}

In contrast, the Alpaca control queries consistently produce zero
fingerprint hits across all distillation settings, indicating that the
observed inheritance is specific to the target semantic domain rather
than caused by general behavioral imitation. All settings with at least
5\% GSM8K data exceed the ownership threshold of 22 responses, suggesting
that \methodname fingerprints can be effectively transferred through
black-box knowledge distillation when a small amount of target-domain
behavior is exposed to the student model.

\FloatBarrier

\end{document}